\documentclass[letterpaper,twocolumn,10pt]{article}
\usepackage{usenix}

\usepackage{tikz}
\usepackage{amsmath}

\usepackage{filecontents}

\usepackage{amsmath,amssymb,amsfonts}
\usepackage{algorithmic}
\usepackage{textcomp}
\usepackage{xcolor}

\usepackage{subfiles}
\usepackage{graphicx}
\usepackage{subcaption}
\usepackage{threeparttable}
\usepackage{booktabs}
\usepackage{multirow} 
\usepackage{makecell}
\usepackage{enumitem}
\usepackage{array}
\usepackage{xurl}

\begin{document}
%-------------------------------------------------------------------------------

%don't want date printed
\date{}

% make title bold and 14 pt font (Latex default is non-bold, 16 pt)
\title{\Large \bf Tanchun: Robust Flow-Based HTTP(S) DDoS Detection via \\Behavioral Attribute Guidance}

%for single author (just remove % characters)
% \author{
% {\rm Your N.\ Here}\\
% Your Institution
% \and
% {\rm Second Name}\\
% Second Institution
% % copy the following lines to add more authors
% % \and
% % {\rm Name}\\
% %Name Institution
% } % end author

\author{
{\rm Qingyang Li}$^{*}$ \quad
{\rm Yunpeng Tan}$^{*}$ \quad
{\rm Mingxin Yang}$^{*}$ \quad
{\rm Yannan Hu}$^{\dagger}$ \quad
{\rm Lei Zhang}$^{\dagger}$ \quad
{\rm Yong Cui}$^{\ddagger}$ \quad \\
{\rm Zongming Guo}$^{*}$ \quad
{\rm Xinggong Zhang}$^{*}$
\\[0.5em]
$^{*}$Peking University \quad
$^{\dagger}$Zhongguancun Laboratory \quad
$^{\ddagger}$Tsinghua University
}

\maketitle

%-------------------------------------------------------------------------------
\begin{abstract}

HTTP(S) DDoS attacks have surged in frequency and scale, posing a significant threat to network security. Flow-based detection is a promising solution for large-scale HTTP(S) DDoS detection due to its decryption-free operation and scalability. However, the coarse-grained numerical statistics in flow data limit accurate attack detection and lack robustness when benign and attack traffic patterns drift over time.

In this paper, we present Tanchun, a drift-aware flow-based HTTP(S) DDoS detection system that relies solely on decryption-free flow data. Specifically, Tanchun groups flows by IP-pair and aggregates them over consecutive time slots into behavior sequences, capturing temporal access patterns to compensate for the limited information in individual flows. It further abstracts these sequences into human-understandable behavioral attributes that capture stable behavioral semantics. Tanchun then aligns behavior sequences with their corresponding behavioral attributes to guide representation learning, enabling robust detection under traffic drift.

We evaluate Tanchun across diverse HTTP(S) DDoS scenarios. Tanchun achieves the best overall performance, with an average F1-score of 0.988, and outperforms the second-best method by 5.3\% and 9.6\% under benign-traffic and attack-traffic drift, respectively. Experiments on real-world traffic traces further demonstrate its practical effectiveness.

\end{abstract}
\section{Introduction}

HTTP(S) DDoS attacks have become a critical threat to modern networks. Unlike volumetric DDoS attacks that primarily exhaust network bandwidth, HTTP(S) DDoS attacks overwhelm application-layer services with massive HTTP or HTTPS requests, consuming server-side resources and degrading service availability. Such attacks continue to grow in both frequency and intensity. For example, Cloudflare reported a 118\% year-over-year increase in HTTP flood attacks in Q1 2025, with rates reaching millions of requests per second~\cite{q1_cloudflare}. Meanwhile, HTTPS-based attacks have become increasingly prevalent, accounting for 63.65\% of application-layer DDoS attacks~\cite{huawei_report}. The growing attack scale and prevalence of encryption make scalable detection increasingly challenging.

Flow-based detection offers a promising solution due to its decryption-free operation and scalability. Flow data, readily available from widely deployed telemetry systems~\cite{netflow_claise2004cisco,Securing_Public_Cloud_mani2025securing,li2016flowradar}, abstracts massive packet-level communications into compact five-tuple-based statistics without decrypting payloads. Unlike commercial web application firewalls (WAFs)~\cite{alibaba}, which require HTTPS decryption for content inspection, flow-based detection avoids decryption overhead and reduces privacy concerns in sensitive services. It also incurs lower processing costs~\cite{hsieh2024netvigil} than traffic-analysis methods that rely on per-packet inspection or feature extraction~\cite{mirsky2018kitsune,whisper,hsieh2024netvigil}, making it better suited for deployment in high-speed networks.

Despite these advantages, leveraging flow data for HTTP(S) DDoS detection remains challenging due to two fundamental issues: \textit{coarse-grained flow statistics} and \textit{traffic drift}.

\textbf{Coarse-Grained Flow Statistics.}
Flow data typically provides only coarse-grained statistics, such as byte and packet counts. Attackers can issue legitimate-looking requests and generate low-volume flows that closely resemble benign traffic~\cite{ALDDoS_server2_praseed2018ddos}, thereby diminishing the discriminative capability of individual flows for detection. This limitation motivates us to aggregate flows sharing the same IP-pair (i.e., a source IP and a destination IP) to capture client-server access patterns from a more holistic perspective. While individual flows can appear indistinguishable from benign traffic, aggregated IP-pair traffic patterns may reveal distinctive characteristics that differentiate attackers from legitimate users~\cite{hypervision,hsieh2024netvigil}.

\textbf{Traffic Drift.}
Although existing IP-pair-level analysis methods have achieved strong performance~\cite{hsieh2024netvigil,hypervision}, their IP-pair feature extraction mechanisms still face substantial challenges in the presence of traffic drift. In real-world networks, traffic distributions continuously evolve due to changes in benign user behaviors and attacker strategies. Existing approaches typically characterize IP-pairs using numerical features, such as flow counts, packet/byte volumes~\cite{hsieh2024netvigil}. However, these numerical features are sensitive to traffic drift and may deviate from the patterns observed in training data, thereby degrading detection performance and robustness~\cite{CADE_yang2021cade}.

To overcome this limitation, we aim to find a way to characterize high-level access behaviors and capture stable behavioral semantics that are more robust to traffic drift. Our key insight is that, while numerical IP-pair features may change over time, high-level access behavior may remain stable and can be described by human-understandable behavioral attributes. For example, HTTP(S) DDoS generated by different automated tools may exhibit different traffic characteristics, yet security operators may characterize their per-bot access behaviors with the same behavioral attributes, such as \textit{"high-intensity"} and \textit{"regular"} access patterns. Therefore, leveraging such stable behavioral attributes to construct semantic representations of IP-pair access behavior provides a promising direction for robust detection under traffic drift.

In this paper, we present Tanchun\footnote{Tanchun is a perceptive girl in the classical novel \textit{Dream of the Red Chamber}, and she recognizes the looming threat within the family.}, a drift-aware flow-based HTTP(S) DDoS detection system that relies solely on decryption-free flow data. Tanchun learns stable semantics of IP-pair access behaviors through human-understandable behavioral attributes. Specifically, it first aggregates individual flows into numerical IP-pair features to characterize client--server access patterns. It then describes each IP-pair's access behavior using human-understandable behavioral attributes and leverages these attributes to guide a deep learning model in capturing high-level behavioral semantics for robust attack detection under traffic drift. However, realizing this goal is non-trivial and requires addressing three main technical challenges (TCs).

\textbf{TC 1: How can flows be aggregated to capture the temporal access behavior of IP-pairs?} Client-server access behavior naturally evolves over time, and may generate multiple flows. However, individual flow records provide limited observations, while naive aggregation over a whole observation window like existing works~\cite{hsieh2024netvigil} removes temporal ordering and obscures behavioral dynamics.

\textbf{TC 2: How can behavioral attributes be defined and leveraged to characterize diverse access behaviors?} Access behaviors exhibit diverse patterns, making it challenging to design behavioral attributes that are sufficiently comprehensive while remaining discriminative. Moreover, mapping numerical IP-pair features to behavioral attributes is challenging due to the lack of explicit quantitative criteria.

\textbf{TC 3: How can the model learn robust behavioral semantics while preserving malicious-benign discrimination?}
Behavioral attributes provide stable semantic abstraction but may lose fine-grained information for distinguishing malicious and benign traffic, while numerical IP-pair features preserve detailed patterns but are sensitive to traffic drift.

To address these technical challenges, we develop Tanchun with three key contributions:

\begin{itemize}[nosep]
    \item \textbf{Behavior Sequence Abstraction.} To address TC 1, it groups flows by IP-pair and aggregates flows within each consecutive time slot to extract diverse statistical features. The slot-level features are then concatenated into behavior sequences, capturing the temporal access dynamics of each IP-pair.
    
    \item \textbf{Behavioral Attribute Derivation.} To address TC 2, it first defines some attribute types that describe IP-pair access behavior from different perspectives and their corresponding quantifiable indicators, and then derives behavioral attributes by mapping numerical behavior sequences to different attribute levels.

    \item \textbf{Semantic-Guided Representation Learning.} To address TC 3, it jointly leverages behavioral attributes and behavior sequences. It learns behavior-aware representations by aligning behavior sequences with their corresponding attributes, encouraging IP-pair representations with the same attributes to be close while preserving fine-grained temporal information for detection.
\end{itemize}

We prototype Tanchun and evaluate it across diverse HTTP(S) DDoS scenarios. Tanchun achieves the best overall performance across public datasets, synthetic complex attack scenarios, and diverse traffic-drift settings, with an average F1-score of 0.988. In particular, Tanchun outperforms the second-best method by 5.3\% and 9.6\% under benign-traffic and attack-traffic drift, respectively. Comprehensive ablation studies and parameter analyses further validate the effectiveness of its key design choices, while evaluations on real-world traffic traces demonstrate its practical detection capability.

\section{Background and Motivation}

% \begin{figure*}[t]
%     \centering

%     \begin{minipage}{0.32\textwidth}
%         \centering
%         \includegraphics[width=0.9\linewidth]{chart/security_flow_tsne.pdf}
%         \caption{HTTP(S) DDoS flows confused with benign flows.}
%         \label{fig:flow_visual}
%     \end{minipage}
%     \hfill
%     \begin{minipage}{0.32\textwidth}
%         \centering
%         \includegraphics[width=0.9\linewidth]{chart/security_moti_benign_attack_distance_cdf.pdf}
%         \caption{Distance between Benign and Attack IP-pair feature.}
%         \label{fig:moti_distance_cdf}
%     \end{minipage}
%     \hfill
%     \begin{minipage}{0.32\textwidth}
%         \centering
%         \includegraphics[width=0.9\linewidth]{chart/security_motivation_attribute_match.pdf}
%         \caption{Matching ratios for four attributes under traffic drift.}
%         \label{fig:attributes_match}
%     \end{minipage}

% \end{figure*}

\begin{figure*}[t]
    \centering

    \begin{minipage}{0.32\textwidth}
        \centering
        \includegraphics[width=0.9\linewidth]{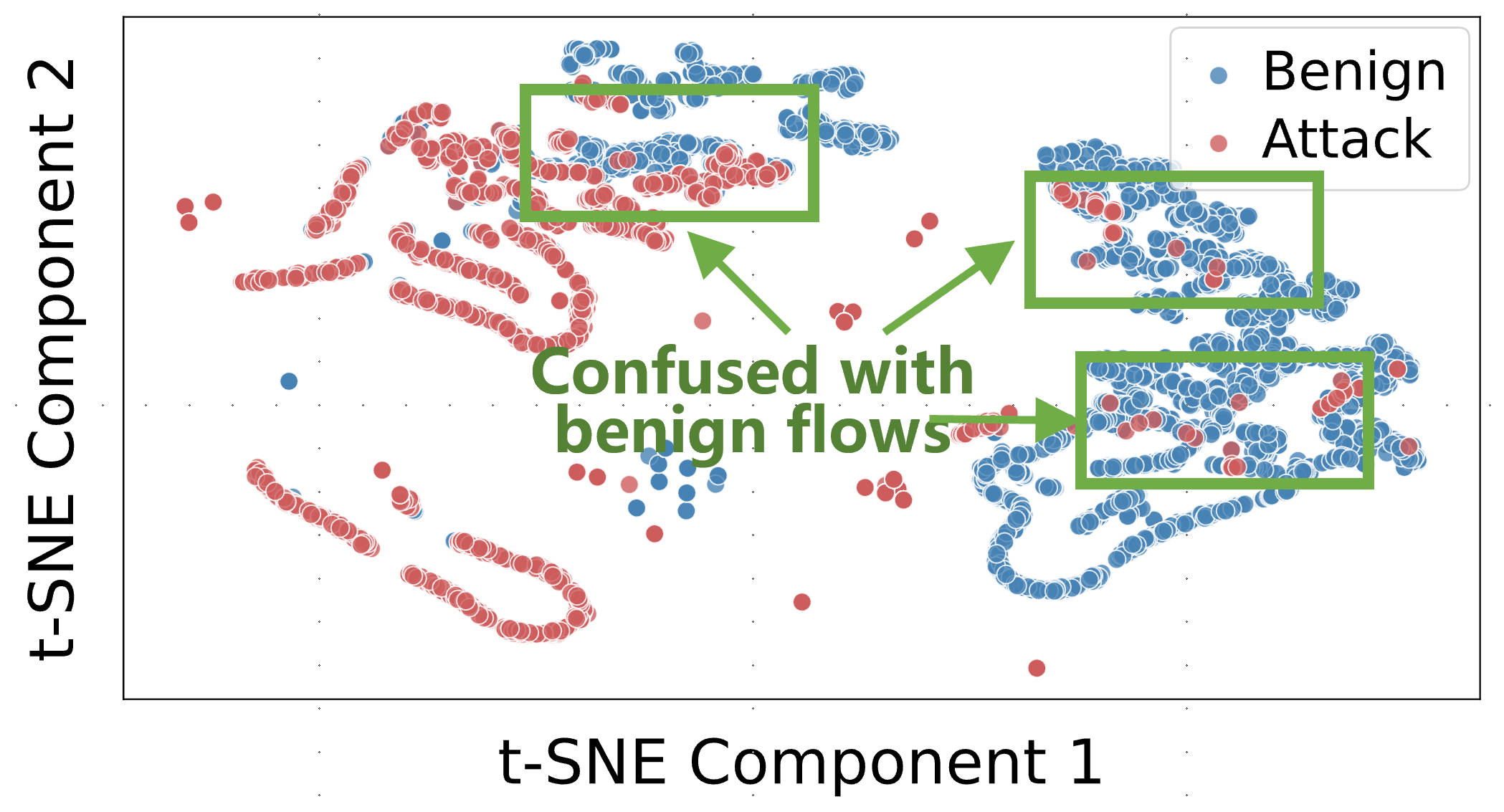}
        \caption{HTTP(S) DDoS flows confused with benign flows.}
        \label{fig:flow_visual}
    \end{minipage}
    \hfill
    \begin{minipage}{0.32\textwidth}
        \centering
        \includegraphics[width=0.9\linewidth]{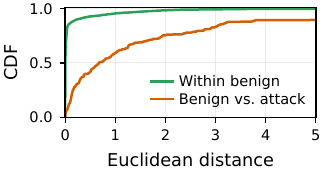}
        \caption{Distances between benign and attack IP-pair features.}
        \label{fig:moti_distance_cdf}
    \end{minipage}
    \hfill
    \begin{minipage}{0.32\textwidth}
        \centering
        \includegraphics[width=0.9\linewidth]{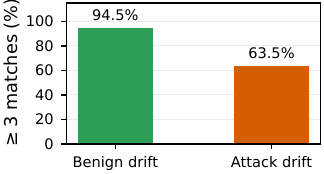}
        \caption{Matching ratios for four attributes under traffic drift.}
        \label{fig:attributes_match}
    \end{minipage}

\end{figure*}

\subsection{Threat Model}
In this paper, we focus on HTTP(S) DDoS attacks, which overwhelm server-side resources by generating excessive HTTP(S) requests rather than exhausting network bandwidth as volumetric DDoS attacks do~\cite{ALDDoS_server2_praseed2018ddos}. These attacks generally fall into two categories~\cite{alddos_black2022overview}: HTTP(S) floods, which consume service resources through massive requests, and slow attacks (e.g., Slowloris), which exhaust connection resources by maintaining numerous persistent HTTP(S) connections.

We assume that attackers control a set of bots to launch HTTP(S) DDoS attacks against target services. Single-bot attacks are treated as a special case and are not distinguished in terminology. Attackers may employ different tools, resulting in diverse request rates, temporal patterns, and resource demands. We assume that attackers attempt to conceal their malicious behavior by keeping individual attack flows low-volume and benign-like, as also observed in our real-world traces (\S~\ref{sub_sec:Stealthy Features}). Highly multiplexed attacks~\cite{HTTP_multiplexing_flood} may expose stronger per-flow characteristics and are therefore outside the primary scope of this work. Attackers may further evade detection by varying the number of bots or reducing per-bot request rates to mimic benign users. We refer to distributed attacks with low per-bot request rates as low-rate attacks~\cite{huawei_report}.

\subsection{Flow-based Detection System}
\label{subsec_flow_system}

HTTP(S) DDoS detection systems should achieve both high accuracy and scalability. Commercial solutions such as WAFs~\cite{alibaba} inspect packet contents but require HTTPS decryption, incurring significant decryption overhead~\cite{huawei_report} and privacy concerns in sensitive domains such as financial services. Traffic-based approaches avoid payload inspection by analyzing traffic patterns~\cite{survey_shen2022machine}; however, many require per-packet processing~\cite{mirsky2018kitsune,whisper}, which remains computationally expensive~\cite{hsieh2024netvigil} and limits scalability when processing aggregate HTTP(S) traffic on high-speed links.

Flow-based detection provides a scalable alternative with two key advantages. First, flow telemetry is widely supported by existing network infrastructures, including routers (e.g., NetFlow~\cite{netflow_claise2004cisco} and IPFIX~\cite{IPFIX_hofstede2014flow}), cloud networks (e.g., NSG and VPC)~\cite{Securing_Public_Cloud_mani2025securing}, and other network monitoring systems~\cite{li2016flowradar,turboflow_sonchack2018turboflow}. Second, flow data enables efficient and privacy-preserving detection. By aggregating massive packets into compact five-tuple (\textit{srcIP}, \textit{dstIP}, \textit{srcPort}, \textit{dstPort}, \textit{protocol}) level statistics, flow data avoids payload inspection and significantly reduces processing overhead compared with per-packet analysis~\cite{hsieh2024netvigil}. 

\textbf{System Model.} Leveraging the scalability of flow-based detection, our system detects HTTP(S) DDoS attacks using flow data. A flow collector deployed at the gateway of the protected server cluster captures inbound traffic and extracts flow records for the detector. The detector analyzes flows targeting the HTTP(S) service ports exposed by the protected servers and identifies attack flows. The detection results can then trigger mitigation mechanisms such as IP blocking and FlowSpec. Since existing flow telemetry systems typically expose only limited fields, including flow start time, packet counts, byte counts, and duration~\cite{netflow_claise2004cisco,IPFIX_hofstede2014flow}, our system relies solely on these commonly available features.

\subsection{Limited Flow Discriminability}
\label{sub_sec:Stealthy Features}

However, flow-level features are often too coarse-grained to accurately characterize HTTP(S) DDoS. To demonstrate this limitation, we analyze five real-world HTTP(S) DDoS traces collected from the production DDoS scrubber of a top-tier ICT vendor~\cite{huawei_12000}, whose authenticity is validated by security operators. Since DDoS scrubbers only retain anomalous traffic for mitigation, benign flows from the same environment are unavailable. Therefore, we use HTTP(S) traces from the MAWI dataset~\cite{mawi_kenjiro2000traffic} (January 2024) as benign references, containing flows generated by diverse users under realistic network conditions. We extract flow features from both attack and benign traces and visualize them using t-SNE.

As shown in Fig.~\ref{fig:flow_visual}, HTTP(S) DDoS and benign flows are highly entangled in the feature space, making them difficult to distinguish based solely on flow features. Similar limitations of flow-level features have also been reported by prior studies~\cite{http_related_rl_feng2020application}. This is mainly because HTTP(S) DDoS attacks can exploit legitimate requests while maintaining low per-flow volume~\cite{ALDDoS_server2_praseed2018ddos}, causing attack flows to resemble benign flows. Therefore, relying solely on individual-flow analysis is insufficient for accurate HTTP(S) DDoS detection.

% \begin{figure}
%     \centering
%     \includegraphics[width=0.8\linewidth]{chart/security_flow_tsne.pdf}
%     \caption{HTTP(S) DDoS flows confused with benign flows.}
%     \label{fig:flow_visual}
% \end{figure}

% The low-volume legitimate requests in HTTP(S) DDoS contribute to its stealthy flow features, as demonstrated in the benchmark dataset CIC-IDS2018~\cite{cic18-sharafaldin2018toward}. We conduct a t-SNE visualization on HTTP(S) benign flows and three types of HTTP(S) DDoS attacks—LOIC, GoldenEye, and Slowloris—using NetFlow-derived features. As shown in Fig.~\ref{fig:flow_visual}, HTTP(S) DDoS flows and benign flows exhibit significant overlap in the feature space, making them difficult to distinguish. Therefore, relying solely on single-flow features is insufficient for accurately identifying HTTP(S) DDoS flows. This observation motivates us to further explore the core characteristics of HTTP(S) DDoS attacks, particularly the malicious request behavior exhibited by bots. Because, unlike benign users, bots are typically script-driven and follow predetermined attack patterns to maximize damage.

\subsection{From Flows to IP-Pair Analysis}
\label{subsec_moti_behavior_analysis}

Although individual flow features of attack and benign traffic may appear similar, their aggregated client–server interaction behaviors can be more distinguishable. Recent studies~\cite{hsieh2024netvigil,hypervision} therefore group flows by IP pairs (i.e., source and destination IPs) and extract aggregated features to characterize access behaviors. For example, NetVigil~\cite{hsieh2024netvigil} uses statistical features such as flow counts and byte volumes, while HyperVision~\cite{hypervision} aggregates small flows as edges for graph-based detection.

\textbf{Benefit of IP-Pair Aggregation.}
IP-pair aggregation provides a natural way to characterize HTTP(S) DDoS behaviors by capturing client–server access patterns beyond individual flows. To examine this effect, we analyze the traces described in \S~\ref{sub_sec:Stealthy Features} by grouping flows into IP-pairs and extracting four features: \textit{flow count}, \textit{byte count}, \textit{packet count}, and the \textit{standard deviation of per-flow byte counts}, following existing studies~\cite{hsieh2024netvigil}. After feature standardization, we calculate distances among benign IP-pairs and between benign and attack IP-pairs. Fig.~\ref{fig:moti_distance_cdf} presents the cumulative distribution functions (CDFs). As shown, benign–attack IP-pair distances are significantly larger than benign IP-pair distances, indicating that IP-pair aggregation reveals more discriminative behavioral patterns. This observation motivates modeling IP-pair access behaviors rather than relying solely on individual flows.

% \begin{figure}
%     \centering
%     \includegraphics[width=0.75\linewidth]{chart/security_moti_benign_attack_distance_cdf.pdf}
%     \caption{Distance between Benign and Attack IP-pair feature.}
%     \label{fig:distance_cdf}
% \end{figure}

\textbf{Challenges Under Traffic Drifts.} However, IP-pair aggregation can be affected by traffic drifts commonly observed in real-world networks. Benign access patterns naturally evolve over time, while attackers continuously adjust their strategies to evade detection, causing numerical IP-pair features to vary across conditions. To quantify this impact, we analyze two benign HTTP(S) traces collected one week apart (January 1 and January 8, 2024)~\cite{mawi_kenjiro2000traffic} and two real-world attack traces representing HTTP and HTTPS floods. We measured the relative changes in IP-pair features, as shown in Table~\ref{tab:mean_relative_drift}.

As shown, both benign and attack traffic evolution can introduce substantial drifts in IP-pair features. The average relative changes reach 34.01\% for benign IP-pairs and 227.16\% for attack IP-pairs, indicating significant variations across traffic conditions. Since existing approaches typically rely on numerical IP-pair features as model inputs, such variations introduce distribution drift between training and detection data, potentially degrading detection robustness~\cite{CADE_yang2021cade}.

\begin{table}[h]
    \centering
    \caption{Relative changes in numerical IP-pair features under benign and attack traffic drift.}
    \label{tab:mean_relative_drift}
    \setlength{\tabcolsep}{3.5pt}
    \renewcommand{\arraystretch}{1.0}
    \resizebox{\columnwidth}{!}{%
        \begin{tabular}{lccccc}
            \toprule
            \textbf{Drift type}
            & \textbf{Flow count}
            & \textbf{Bytes}
            & \textbf{Packets}
            & \textbf{Flow std}
            & \textbf{Mean} \\
            \midrule
            Benign drift
            & 69.77\%
            & 16.83\%
            & 14.31\%
            & 35.13\%
            & \textbf{34.01\%} \\
            
            Attack drift
            & 344.22\%
            & 459.21\%
            & 6.84\%
            & 98.39\%
            & \textbf{227.16\%} \\
            \bottomrule
        \end{tabular}%
    }
\end{table}

\subsection{Behavioral Attributes for Stable Semantics}
\label{moti_behavior_attri}

To mitigate the drift issue, we aim to derive a robust representation of IP-pair access behaviors beyond numerical features.

\textbf{Our Intuition.}
Although traffic drifts may alter numerical IP-pair features, the underlying access behaviors can remain consistent. For example, HTTP(S) DDoS attacks launched with different automated tools may exhibit diverse traffic volumes and temporal patterns. However, they may still exhibit consistent malicious behaviors that security experts can describe using human-understandable attributes, such as \textit{"high-intensity"} and \textit{"regular"} per-bot access patterns. This motivates us to explore behavioral attributes as a more stable representation of IP-pair access behaviors beyond numerical features. Inspired by attribute-based representations in computer vision, where visual attributes have been leveraged to improve generalization to unseen image classes~\cite{cv_attribute_xu2020attribute,cv_attribute_al2015transfer}, we investigate behavioral attributes for robust HTTP(S) DDoS detection under traffic drifts.

\textbf{Observation.}
To preliminarily evaluate the stability of behavioral attributes under traffic drifts, we transform the four IP-pair features introduced in \S~\ref{subsec_moti_behavior_analysis} into coarse-grained attributes. Specifically, we derive three \textit{"access intensity"} attributes from flow, packet, and byte volumes, and one \textit{"access diversity"} attribute from flow-size variations. Each feature is mapped to one of three attribute levels (high, medium, and low) using percentile-based thresholds obtained from benign traffic, and the same thresholds are consistently applied across all traces.
As shown in Fig.~\ref{fig:attributes_match}, these attributes remain robust under traffic drifts: 94.5\% of benign comparisons and 63.5\% of attack comparisons preserve at least three consistent attributes. Even with substantial changes in attack vectors (e.g., HTTP floods to HTTPS floods), the attributes maintain meaningful consistency, demonstrating their potential for robust IP-pair behavior modeling under traffic drifts.

% \begin{figure}
%     \centering
%     \includegraphics[width=0.7\linewidth]{chart/security_motivation_attribute_match.pdf}
%     \caption{Matching ratios for four attributes under traffic drift.}
%     \label{fig:attributes_match}
% \end{figure}

\begin{figure*}
    \centering
    \includegraphics[width=0.8\linewidth]{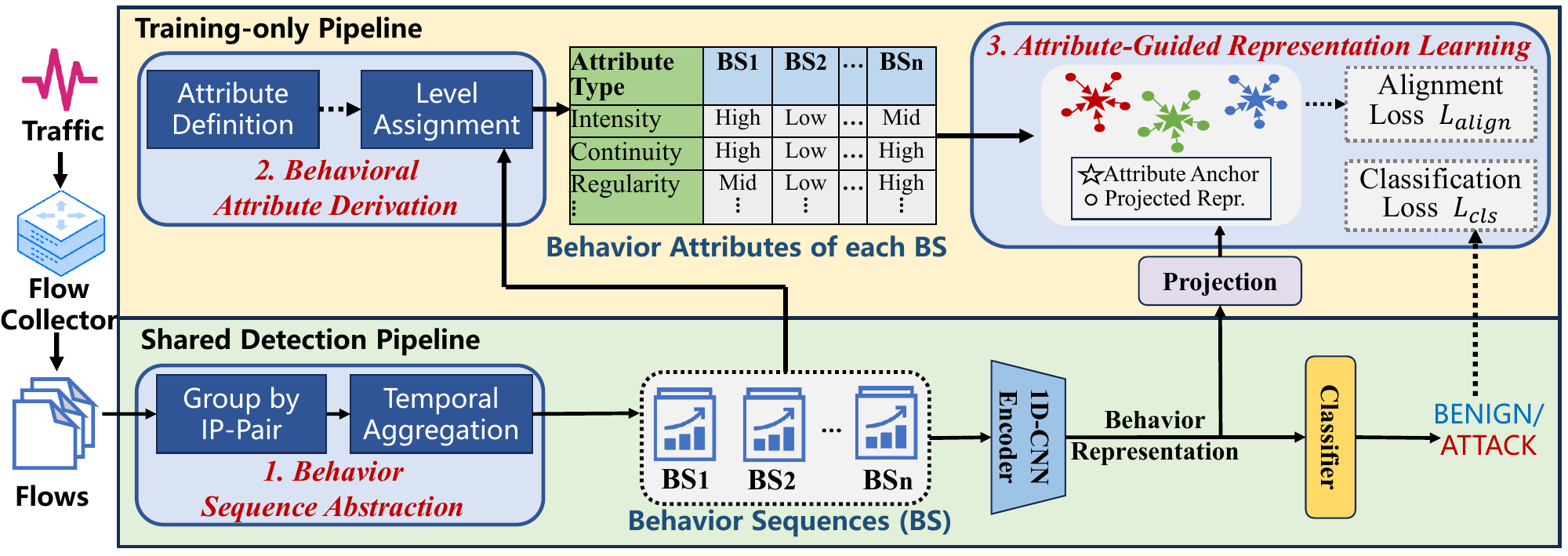}
    \caption{The overall architecture and workflow of Tanchun.}
    \label{fig:method_overview}
\end{figure*}

\subsection{Key Insight and Technical Challenges}
\label{sub_sec:insights}

Based on the above observations, we derive the following insight: robust HTTP(S) DDoS detection under traffic drift requires modeling the stable semantics of IP-pair access behaviors rather than relying solely on numerical features. Behavioral attributes offer a promising representation, as they may capture stable behavioral characteristics despite variations in numerical features under traffic drift.

However, incorporating this insight into a practical system is non-trivial, and we face three key technical challenges:

\textbf{TC 1: How can flows be aggregated to capture the temporal access behavior of IP-pairs?} Client–server access naturally evolves over time and may generate multiple flows with diverse temporal characteristics. Existing approaches typically aggregate flows within an observation window to extract statistical features~\cite{hsieh2024netvigil,hypervision}. However, such aggregation removes the temporal ordering among flows and may obscure important behavioral dynamics, limiting the ability to capture the evolving access behaviors of IP pairs.

\textbf{TC 2: How can behavioral attributes be defined and leveraged to characterize diverse access behaviors?} Access behaviors exhibit substantial diversity across users and attack strategies, making it challenging to design attributes that provide sufficient coverage while preserving discriminative capability. Poorly designed attributes may miss critical behavioral patterns or fail to distinguish malicious access from benign activities. Although our preliminary analysis (\S~\ref{moti_behavior_attri}) demonstrates the potential of behavioral attributes under traffic drifts, it considers only a limited set of attributes and cannot capture the full diversity of real-world behaviors. Moreover, behavioral attributes are high-level semantic abstractions without predefined quantitative mappings, making their derivation from numerical IP-pair features non-trivial.

\textbf{TC 3: How can the model learn robust behavioral semantics while preserving malicious-benign discrimination?}
Behavioral attributes abstract IP-pair access behaviors and are therefore more robust to traffic drifts. However, this abstraction may discard fine-grained information that is critical for accurate classification. In contrast, numerical IP-pair features retain detailed traffic characteristics that facilitate malicious-benign discrimination, but are more susceptible to traffic drifts. The key challenge is thus to learn representations that capture stable behavioral semantics while preserving the discriminative information encoded in numerical features.

% \subfile{sections/system overview and Treat Model.tex}

\section{The Design of Tanchun}

\subsection{Overview}

In this paper, we present Tanchun, a flow-based HTTP(S) DDoS detection system that learns robust semantics of IP-pair access behaviors through human-understandable behavioral attributes. The overall architecture and workflow of Tanchun are illustrated in Fig.~\ref{fig:method_overview}. Specifically, we design the following modules to address the challenges introduced in \S\ref{sub_sec:insights}:

\textbf{Behavior Sequence Abstraction (\S\ref{subsec_sequence_Abstraction}).} This module is designed \textbf{for TC 1} to transform individual flow records into multivariate time series, termed \textit{behavior sequences}, which characterize the temporal patterns of IP-pair access behavior. Specifically, flows associated with each IP-pair are grouped into consecutive time slots, from which a diverse set of statistical features is extracted. The resulting slot-level feature vectors are then chronologically concatenated to form a multivariate behavior sequence for model training and detection.

% \textbf{Indicator-Based Behavioral Attribution.} This module is designed \textbf{for TC 2}. It first defines human-understandable behavioral attributes that describes access pattern in different perspectives and their corresponding quantifiable indicators, and then derives attribute profiles by mapping numerical behavior sequences to attribute levels.

% \textbf{Attribute-Guided Representation Learning.} This module is designed \textbf{for TC 3}. It jointly leverages behavioral attributes and numerical IP-pair Specifically, it learns behavior-aware representations by aligning numerical features with behavioral semantics, clustering IP-pairs with similar behavioral attributes, and preserving discriminative information for attack detection.

\textbf{Behavioral Attribute Derivation (\S~\ref{subsec_BAD}).}
This module is designed for \textbf{TC 2} to describe behavior sequences using human-understandable behavioral attributes. Specifically, it defines seven attribute types based on domain knowledge of HTTP(S) DDoS, providing a comprehensive characterization of IP-pair access behaviors while capturing key differences between benign and attack IP-pairs. It further introduces quantitative indicators for each attribute type, derived from behavior sequences and mapped into three levels, i.e., High, Medium, and Low. Finally, each behavioral attribute is represented as a combination of an attribute type and its corresponding level.

\textbf{Attribute-Guided Representation Learning (\S~\ref{subsec_ARL}).} 
This module is designed for \textbf{TC 3} by leveraging behavioral attributes as semantic supervision to refine behavior sequence representations. Specifically, it designs two complementary learning tasks: \textit{attribute alignment} and \textit{binary classification}. The former aligns behavior sequence representations with their corresponding behavioral attributes, enabling the model to learn transferable representations of typical behavioral patterns. The latter preserves fine-grained discriminative information for attack detection. By jointly optimizing these tasks, Tanchun learns robust behavior representations with improved robustness and detection accuracy.

\subsection{Behavior Sequence Abstraction}
\label{subsec_sequence_Abstraction}

\begin{figure}
    \centering
    \includegraphics[width=0.99\linewidth]{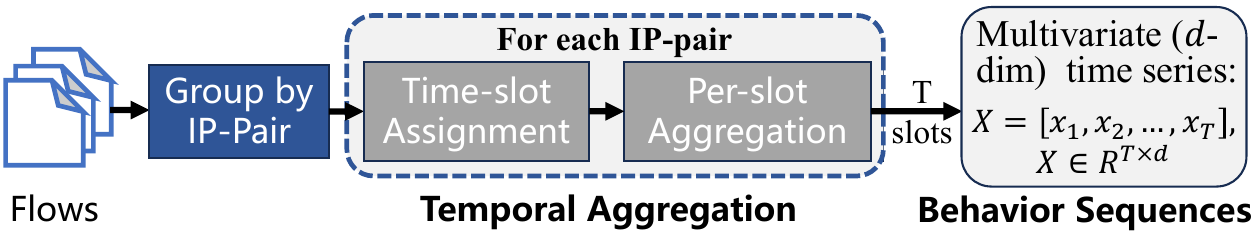}
    \caption{Pipeline of Behavior Sequence Abstraction}
    \label{fig:bsa}
\end{figure}

We propose \textbf{Behavior Sequence Abstraction (BSA)} to transform flow data into time series that capture the temporal access behaviors of IP-pairs. Our key idea is that individual flows only provide instantaneous summaries of client–server interactions, while aggregating flows over consecutive time slots can reveal the evolution of access behaviors over time.

BSA first groups flows by IP pairs and then performs \textbf{Temporal Aggregation} to convert the flows associated with each IP-pair into a multivariate time series, referred to as a \textit{behavior sequence}. The pipeline of BSA is illustrated in Figure~\ref{fig:bsa}. Specifically, Temporal Aggregation consists of two steps: time-slot assignment and per-slot aggregation.

\textbf{Step 1. Time-slot Assignment.}
The start timestamp of each flow indicates when the corresponding communication begins. Therefore, BSA uses flow start times to organize flows along the temporal dimension. Specifically, for each IP-pair, we divide the observation period into consecutive $\Delta t$-second time slots and assign flows whose start times fall within the same time slot to the corresponding interval.

\textbf{Step 2. Per-slot Aggregation.}
Flows within the same time slot provide aggregated information about the access
behavior of an IP-pair during that interval. Therefore, BSA aggregates the
flows in each time slot into a $d$-dimensional feature vector using the
statistical features listed in Table~\ref{tab:agg_features}. Specifically, the
number of flows $N$ captures the access frequency, while $\sum
\textit{Packets}$ and $\sum \textit{Bytes}$ capture the overall traffic volume.
The \textit{mean, std, min, and max} values capture the size distribution and
heterogeneity of individual flows. For time slots containing no flows, all
aggregated features are set to zero.

After temporal aggregation, BSA concatenates $L$ consecutive time-slot
features into a multivariate time series $X_i \in \mathbb{R}^{L \times d}$,
forming the $i$-th behavior sequence. Such behavior sequences characterize
the temporal evolution of IP-pair access.

\begin{table}[t]
    \centering
    \renewcommand{\arraystretch}{0.9} % 控制行距
    \small
    \caption{The aggregated features in each time slot.}
    \label{tab:agg_features}
    \begin{tabular}{@{}cc@{}}
        \toprule
        \textbf{Features} & \textbf{Description} \\
        \midrule
        $N$ & Number of flows \\
        $\sum \textit{Packets}$, $\sum \textit{Bytes}$ & Total number of packets/bytes \\
        $\mu_{\text{Pkt}}$, $\mu_{\text{Byt}}$ & Mean packets/bytes per flow \\
        $\sigma_{\text{Pkt}}$, $\sigma_{\text{Byt}}$ & Std. dev. of packets/bytes per flow \\
        $Min_{\text{Pkt}}$, $Min_{\text{Byt}}$ & Minimum packets/bytes per flow \\
        $Max_{\text{Pkt}}$, $Max_{\text{Byt}}$ & Maximum packets/bytes per flow \\
        \bottomrule
    \end{tabular}
\end{table}

% These behavior sequences will be used for training and detection, and the details will be introduced in \S~\ref{subsec_Put_Them_Together}.

% \begin{figure}
%     \centering
%     \includegraphics[width=0.99\linewidth]{images/behavior sequence generation_new.pdf}
%     \caption{The pipeline for generating behavior sequences from flow data.}
%     \label{fig:behavior_sequence}
% \end{figure}

\subsection{Behavioral Attribute Derivation}
\label{subsec_BAD}

\begin{figure}[t]
    \centering
    \includegraphics[width=0.95\linewidth]{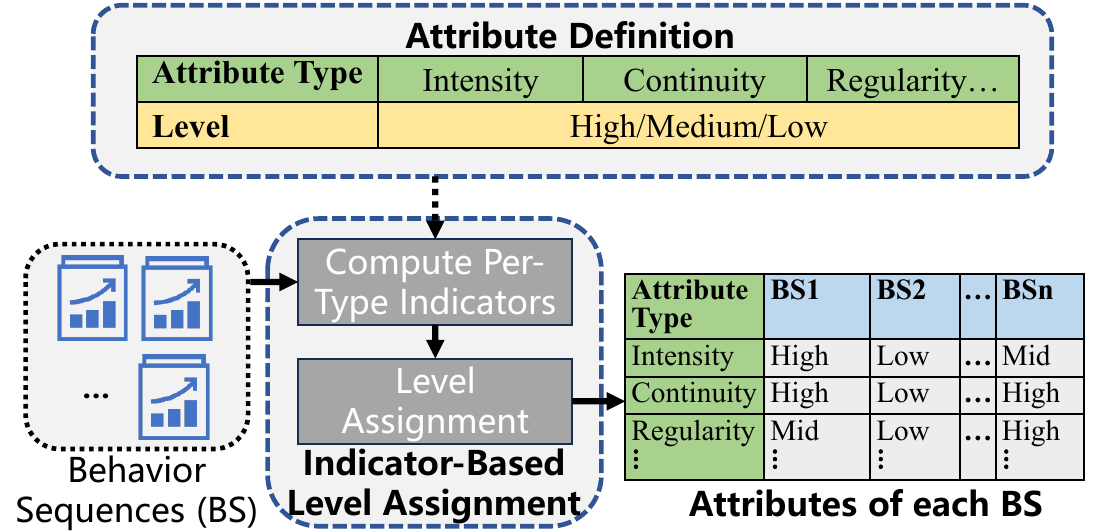}
    \caption{Pipeline of Behavioral Attribute Derivation.}
    \label{fig:BAD}
\end{figure}

Next, we propose \textbf{Behavioral Attribute Derivation (BAD)} to derive human-understandable behavioral attributes from behavior sequences. As illustrated in Fig.~\ref{fig:BAD}, BAD first defines behavioral attributes by specifying attribute types and their corresponding levels for describing IP-pair access behaviors. It then employs an indicator-based level assignment strategy, where quantitative indicators serve as measurable proxies to map behavior sequences to the corresponding levels.

\textbf{Behavior Attribute Definition.}
A behavioral attribute is defined as a combination of an \textit{attribute type} and its corresponding \textit{level}. The attribute type captures a specific aspect of IP-pair access behavior, while the level represents the degree of this characteristic using three discrete levels, i.e., \textit{Low}, \textit{Medium}, and \textit{High}. For example, an IP-pair exhibiting high access intensity is described as \textit{High Access Intensity}.

\begin{table*}[t]
\centering
\caption{Attribute types and corresponding indicators.}
\label{tab:behavioral_attributes}

\small
\renewcommand{\arraystretch}{0.8}

\begin{tabular}{p{2.3cm} p{3.5cm} p{6.2cm} p{4cm}}
\toprule
\textbf{\centering Perspective\arraybackslash} &
\textbf{\centering Attribute Type\arraybackslash} &
\textbf{\centering Interpretation\arraybackslash} &
\textbf{\centering Quantitative  Indicator\arraybackslash} \\

\midrule

Access Frequency
& Access Intensity
& Whether a client accesses the server intensively
& Average flows per active slot
\\

\midrule

\multirow{2}{*}{\makecell{Access\\Activity}}
& Activity Persistence
& Whether access persists throughout the observation window
& Ratio of active slots over $L$-slots
\\

& Activity Continuity
& Whether active slots form continuous segments or are interrupted by idle slots
& Ratio of longest continuous active slots over $L$-slots
\\

\midrule

\multirow{2}{*}{\makecell{Temporal\\Dynamics}}
& Temporal Regularity
& Whether access follows regular temporal patterns
& $S(\cdot)$ of active-slot intervals
\\

& Access Rate Stability
& Whether access rate remains stable
& $S(\cdot)$ of flow counts in active slot
\\

\midrule

\multirow{2}{*}{\makecell{Flow\\Consistency}}
& Packet-Count Consistency
& Whether packet counts are consistent across flows
& $S(\cdot)$ of per-flow pkt counts
\\

& Byte-Volume Consistency
& Whether byte volumes are consistent across flows
& $S(\cdot)$ of per-flow byte counts
\\

\bottomrule
\end{tabular}

\end{table*}

\textbf{Attribute Types}. We design the attribute types from four complementary perspectives motivated by potential differences in access behavior between benign users and HTTP(S) DDoS bots~\cite{huawei_report,http_related_seq_liao2015application,http_related_user_behavior_singh2018user}. These perspectives describe IP-pair behavior in terms of \textit{Access Frequency}, \textit{Access Activity}, \textit{Temporal Dynamics}, and \textit{Flow Consistency}. Under each perspective, we define attribute types to capture specific access behavior.

\textit{(1) Access Frequency}. Benign users typically request only the resources needed for their current service demands, whereas bots may deliberately generate large numbers of requests to amplify attack impact~\cite{huawei_report}. Under this perspective, we define \textit{Access Intensity} to characterize whether a client accesses the server with high intensity.

\textit{(2) Access Activity}. Benign requests are often interspersed with user interactions such as reading, resulting in intermittent activity. In contrast, automated bots often sustain request generation during attacks, leading to longer active periods with fewer idle interruptions~\cite{http_related_seq_liao2015application}. We therefore define \textit{Activity Persistence} and \textit{Activity Continuity} to capture these two aspects of access activity.

\textit{(3) Temporal Dynamics}. Benign access is driven by changing user demands and interactions, resulting in variations in request timing and access rates. In contrast, script-controlled bots may follow more predefined temporal patterns~\cite{http_related_user_behavior_singh2018user}. We therefore define two attribute types, \textit{Temporal Regularity} and \textit{Access Rate Stability}, to characterize the regularity of request timing and the stability of access rates over time.

\textit{(4) Flow Consistency}. Benign users may access diverse resources according to individual interests, resulting in heterogeneous flows. In contrast, bots may repeatedly request the same or similar resources according to attack configurations~\cite{http_related_user_behavior_singh2018user}, which can lead to more consistent packet-count and byte-volume characteristics across flows. We therefore define \textit{Packet-Count Consistency} and \textit{Byte-Volume Consistency} to capture these two aspects of flow-level consistency.

Overall, these four perspectives yield seven attribute types, as summarized in Table~\ref{tab:behavioral_attributes}.

% Overall, these four perspectives yield seven attribute types, as summarized in Table~\ref{tab:behavioral_attributes}. For each attribute type, we define three levels, i.e., Low, Medium, and High, to indicate the degree of the corresponding behavioral characteristic. A behavioral attribute is then represented by an attribute type together with its assigned level. For example, an IP-pair with a high access frequency is described as \textit{High Access Intensity}.

\textbf{Quantitative Indicator.}
Given the defined attribute types, we next determine their corresponding levels for each behavior sequence. Since these attribute types represent high-level behavioral abstractions and cannot be directly derived from behavior sequences, we associate each type with a quantitative indicator as its measurable proxy.

The detailed computation of each indicator is summarized in Table~\ref{tab:behavioral_attributes}. We refer to a time slot containing at least one flow for an IP-pair as an \textit{active slot}. \textit{Access Intensity} is measured by the average flow count within active slots, reflecting the short-term access intensity during active periods. Under the \textit{Access Activity} perspective, we use ratio-based indicators over the $L$-slot behavior sequence to capture the temporal coverage and continuity of active behavior. For attribute types under the \textit{Temporal Dynamics} and \textit{Flow Consistency} perspectives, we quantify variation using the coefficient of variation (CV) and convert it into a bounded stability score $S$:

\begin{equation}
\label{eq:stability_score}
CV = \frac{\sigma}{\mu}, \qquad
S = \frac{1}{1+CV}, \quad S \in [0,1],
\end{equation}
where $\mu$ and $\sigma$ denote the mean and standard deviation of the corresponding observations. A larger $S$ indicates more regular, stable, or consistent behavior.

\textbf{Level Assignment.} After computing the indicator values, we map each value to three levels using two thresholds, $\theta_a^{\mathrm{low}}$ and $\theta_a^{\mathrm{high}}$. Specifically, the indicator value $r_{i,a}$ of the $a$-th attribute type for behavior sequence $i$ is assigned to Low, Medium, or High when $r_{i,a}\leq\theta_a^{\mathrm{low}}$, $\theta_a^{\mathrm{low}}<r_{i,a}\leq\theta_a^{\mathrm{high}}$, or $r_{i,a}>\theta_a^{\mathrm{high}}$, respectively. Ratio-based indicators and stability scores $S(\cdot)$ are bounded between $0$ and $1$, allowing fixed thresholds to be applied on a common scale. In contrast, \textit{Access Intensity}, measured by flows per active slot, has no natural upper bound. We therefore derive its thresholds from the benign training distribution using percentile-based criteria, providing a fixed reference for mapping access intensity into different semantic levels rather than tracking the instantaneous benign distribution.

% \textbf{Behavioral Attribution.}
% IBA performs behavioral attribution for each behavior sequence based on the definition of behavioral attributes. Specifically, IBA first calculates the indicator value associated with each behavioral dimension from the given behavior sequence. Each indicator value is then compared with predefined criteria to determine its corresponding level. Since most of the designed indicators are normalized measurements, such as ratios and stability scores, their values are bounded within interpretable ranges, making the determination of level assignment criteria relatively objective and facilitating consistent attribute derivation. Finally, IBA combines each behavioral dimension with its assigned level to generate a set of human-understandable behavioral attributes for each behavior sequence. 

\subsection{Attribute-Guided Representation Learning}

\label{subsec_ARL}

% \zhang{
% 1, Pipeline intro. Explain the reason of semantic-aligned contrastive learning

% 2, Descriptions - [Text Encoder] - semantic embeddings  -|
% Flow sequence - [Sequence Encoder] - flow embeddings -|- [Semantic-Aligned Contrastive Leaning] - benign/malicious.
% }

% After the above process, we obtained the behavior sequences that characterize the traffic pattern of user requests, along with their corresponding descriptions. To enhance the representation of traffic patterns with the semantics in the descriptions, we employed contrastive learning for representation fusion. Through this method, we established a correspondence between the features of behavior sequences and the malicious request behaviors that can be described in natural language. We refer to this method as \textit{behavior-description contrastive learning}, with the architecture shown in Figure~\ref{fig:cl}. 

\begin{figure}
    \centering
    \includegraphics[width=0.95\linewidth]{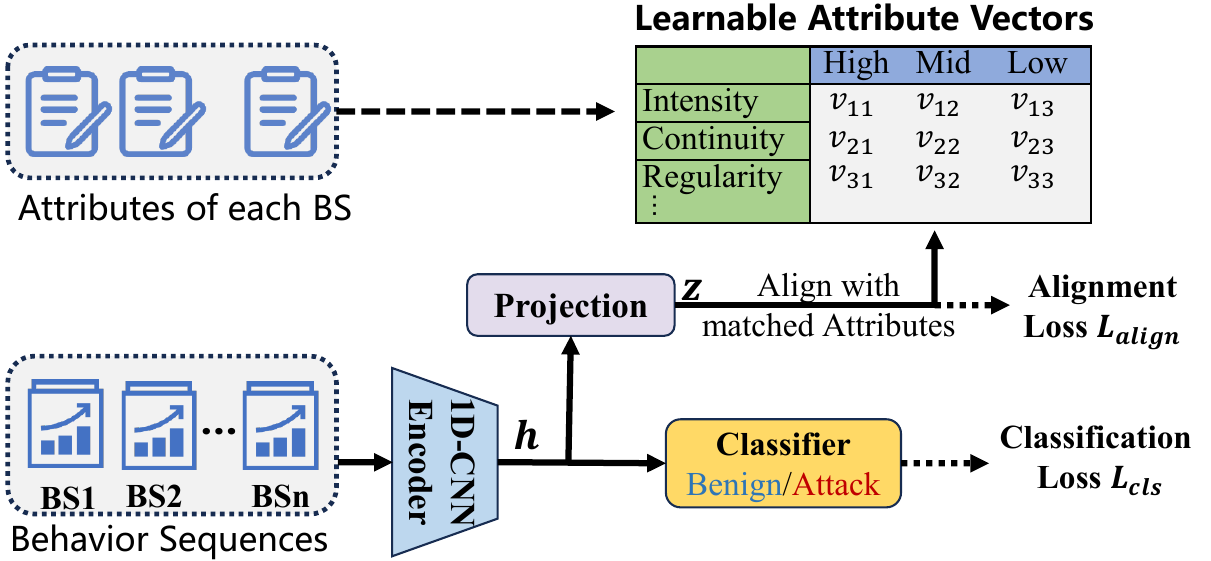}
    \caption{Attribute-Guided Representation Learning.}
    \label{fig:ARL}
\end{figure}

After obtaining the behavior sequences and their corresponding behavioral attributes, we develop an \textbf{attribute-guided representation learning (ARL)} framework to learn representations of behavior sequences under the guidance of behavioral attributes. As illustrated in Fig.~\ref{fig:ARL}, ARL maps behavior sequences and their corresponding attributes into a shared latent space and encourages their representations to be aligned. To this end, we introduce two complementary learning tasks.

\textbf{Attribute Alignment Task.}
Let $\mathcal{X}$ denote the set of behavior sequences. For each behavior
sequence $X_i \in \mathcal{X}$, we employ a two-layer
1D-CNN~\cite{1D-CNN_ige2024state} encoder $E_b(\cdot)$ to extract its behavior
representation $\mathbf{h}_i = E_b(X_i)$. We adopt a 1D-CNN due to its
lightweight architecture and effectiveness in modeling sequential patterns.
For each attribute type $a \in \{1,\ldots,A\}$ and level $k \in \{1,\ldots,K\}$,
we introduce a learnable latent vector $\mathbf{v}_{a,k}$ as the semantic anchor. These latent vectors are randomly
initialized and optimized from scratch during training, allowing the model to
learn task-specific latent space for different behavioral attributes.
To align behavior representations with these attribute anchors, we further
apply a learnable linear projection to $\mathbf{h}_i$, yielding the projected
representation $\mathbf{z}_i$ in the shared latent space. Both $\mathbf{z}_i$
and $\mathbf{v}_{a,k}$ are $\ell_2$-normalized.

Since each behavior sequence is associated with one level for every attribute type, we perform contrastive learning independently across attribute types. For a given type, the matched level is treated as positive, while the remaining levels serve as negatives. Let $c_{i,a} \in {1,\ldots,K}$ denote the matched level of the $i$-th behavior sequence for the $a$-th attribute type. The attribute alignment loss is then defined as
\begin{equation}
\mathcal{L}_{\mathrm{align}}
=
-\frac{1}{|\mathcal{X}|A}
\sum_{i \in \mathcal{X}}
\sum_{a=1}^{A}
\log
\frac{
\exp\left(
\mathbf{z}_i^\top \mathbf{v}_{a,c_{i,a}}/\tau
\right)
}{
\sum_{k=1}^{K}
\exp\left(
\mathbf{z}_i^\top \mathbf{v}_{a,k}/\tau
\right)
},
\end{equation}
where $\tau$ denotes the temperature parameter.

\textbf{Binary Classification Task.}
Since the ultimate goal of our framework is HTTP(S) DDoS detection, we introduce
a classification branch that feeds the behavior representation $\mathbf{h}_i$
into a linear classifier $C(\cdot)$ to predict the attack probability:
$\hat{y}_i = C(\mathbf{h}_i),$
where $\hat{y}_i$ denotes the probability that the $i$-th behavior sequence is
malicious. We use $\mathbf{h}_i$ for classification because it preserves
fine-grained temporal patterns while being jointly optimized by the attribute
alignment task. We adopt binary cross-entropy as the classification loss:

\begin{equation}
\mathcal{L}_{\mathrm{cls}}
=
-\frac{1}{|\mathcal{X}|}
\sum_{i\in\mathcal{X}}
\left[
y_i\log(\hat{y}_i)
+
(1-y_i)\log(1-\hat{y}_i)
\right],
\end{equation}
where $y_i\in\{0,1\}$ denotes the ground-truth of the $i$-th behavior
sequence, indicating benign and malicious,
respectively.

\textbf{Overall Loss Function.}
Finally, we combine the classification and attribute alignment losses as
\begin{equation}
    \mathcal{L}
    =
    \mathcal{L}_{\mathrm{cls}}
    +
    \lambda \mathcal{L}_{\mathrm{align}},
\end{equation}
where $\lambda \in [0,1]$ controls the contribution of the attribute alignment task. During training, the 1D-CNN encoder $E_b(\cdot)$, the linear projection layer, the attribute latent vectors $\mathbf{v}_{a,k}$, and the classifier $C(\cdot)$ are jointly optimized in an end-to-end manner. Through this joint optimization, the model learns to distinguish malicious behaviors while also capturing representative behavioral semantics under attribute guidance. As a result, the learned behavior representations become more transferable.

\subsection{End-to-End Workflow}
\label{subsec_Put_Them_Together}

In this subsection, we describe the end-to-end workflow of Tanchun during
training and detection, as illustrated in Fig.~\ref{fig:method_overview}.

\textbf{Training Phase.}
During training, BSA constructs behavior sequences from training flows using
a sliding window of $L$ time slots with a stride of one time slot. BAD then
derives the corresponding behavioral attributes. The behavior sequences and
their attributes are fed into ARL, which jointly optimizes the model parameters
by minimizing the overall objective $\mathcal{L}$.

\textbf{Detection Phase.}
During detection, BSA constructs behavior sequences from incoming flows using
a truncated window of $L$ time slots to ensure complete flow coverage. Each
sequence $X_i$ is then fed into the 1D-CNN encoder $E_b(\cdot)$ and classifier
$C(\cdot)$ to obtain the attack probability
$\hat{y}_i = C(E_b(X_i))$. A sequence is classified as malicious if
$\hat{y}_i > 0.5$. Notably,
BAD and attribute alignment are only used during training and therefore
introduce no additional inference overhead.

\section{Evaluation}
\subsection{Settings}
\label{eva_settings}

\begin{table*}[t]
\centering
\caption{Overall Performance Comparison with Baselines.}
\label{tab:overall}

\renewcommand{\arraystretch}{0.3}
\setlength{\tabcolsep}{6pt}

\begin{tabular}{
>{\centering\arraybackslash}p{1.6cm}
|
c
|
ccc
|
cc
|
cc
|
c
}

\toprule

\textbf{Methods} & \textbf{Metrics}
& \multicolumn{3}{c}{\textbf{Public Datasets}}
& \multicolumn{2}{c}{\textbf{Complex Attacks}}
& \multicolumn{2}{c}{\textbf{Traffic Drift}}
& \textbf{Overall}
\\

\cmidrule(lr){3-5}
\cmidrule(lr){6-7}
\cmidrule(lr){8-9}

&
& IDS-17 & IDS-18 & IOT-23
& LRHF & Multi-Vec
& Benign Drift & Attack Drift
&
\\

\midrule

\multirow{2}{*}{NetBeacon}
& ACC
& 0.980 & 0.979 & \textbf{0.995}
& 0.995 & 0.942
& 0.986 & 0.553
& 0.919
\\

& F1
& 0.978 & 0.970 & \textbf{0.995}
& 0.997 & 0.956
& 0.932 & 0.075
& 0.843
\\

\addlinespace[0.15em]
\cmidrule(lr){3-9}

\multirow{2}{*}{FSC}
& ACC
& 0.602 & 0.565 & 0.676
& 0.401 & 0.442
& 0.301 & 0.498
& 0.498
\\

& F1
& 0.717 & 0.638 & 0.568
& 0.393 & 0.403
& 0.299 & 0.498
& 0.502
\\

\addlinespace[0.15em]
\cmidrule(lr){3-9}

\multirow{2}{*}{Trident}
& ACC
& 0.582 & 0.593 & 0.682
& 0.721 & 0.430
& 0.412 & 0.558
& 0.568
\\

& F1
& 0.720 & 0.634 & 0.782
& 0.813 & 0.531
& 0.329 & 0.601
& 0.630
\\

\addlinespace[0.15em]
\cmidrule(lr){3-9}

\multirow{2}{*}{NetVigil}
& ACC
& 0.950 & 0.911 & 0.945
& 0.832 & 0.913
& 0.171 & 0.902
& 0.803
\\

& F1
& 0.955 & 0.908 & 0.935
& 0.878 & 0.936
& 0.271 & 0.813
& 0.814
\\

\addlinespace[0.15em]
\cmidrule(lr){3-9}

\multirow{2}{*}{HyperVision}
& ACC
& 0.965 & 0.804 & 0.696
& 0.603 & 0.639
& 0.223 & 0.746
& 0.668
\\

& F1
& 0.967 & 0.799 & 0.771
& 0.752 & 0.780
& 0.272 & 0.763
& 0.729
\\

\addlinespace[0.15em]
\cmidrule(lr){3-9}

\multirow{2}{*}{BS-CNN}
& ACC
& 0.999 & 0.999 & 0.995
& 0.999 & 0.988
& 0.972 & 0.963
& 0.988
\\

& F1
& 0.999 & 0.999 & 0.995
& 0.999 & 0.990
& 0.902 & 0.871
& 0.965
\\

\addlinespace[0.15em]
\midrule

\multirow{2}{*}{\textbf{Tanchun}}
& ACC
& \textbf{0.999} & \textbf{0.999} & 0.992
& \textbf{0.999} & \textbf{0.989}
& \textbf{0.996} & \textbf{0.984}
& \textbf{0.994}
\\

& F1
& \textbf{0.999} & \textbf{0.999} & 0.992
& \textbf{0.999} & \textbf{0.991}
& \textbf{0.982} & \textbf{0.955}
& \textbf{0.988}
\\

\bottomrule

\end{tabular}
\label{tab:overall}
\end{table*}

\textbf{Implementation}. Tanchun is implemented in approximately 1,500 lines of Python 3.10.13 code, with the deep learning model built using \textit{PyTorch 2.2.1} and \textit{CUDA 11.8}. For the six bounded indicators in $[0,1]$, we use thresholds of 0.2 and 0.8, while the unbounded \textit{Access Intensity} uses the 20th and 80th percentiles of benign training traffic. These thresholds separate the lower and upper tails from the central region, corresponding to Low, Medium, and High levels, respectively. We set the time-slot length $\Delta t$ to 1 second and the behavior-sequence length $L$ to 15. The contrastive temperature and attribute alignment weight are set to $\tau=0.1$ and $\lambda=0.2$, respectively. We train the model with AdamW using a learning rate of $10^{-4}$, a batch size of 64, and 32 epochs.

\textbf{Testbed}. We conduct all experiments on a Linux server running Ubuntu 24.04.1 LTS with an AMD Ryzen 9 7950X 32-core CPU, an NVIDIA GeForce RTX 4090D GPU with 24 GB GPU memory, and 128 GB RAM.

\textbf{Baselines.}
According to our system model~\S\ref{subsec_flow_system}, we compare Tanchun with six flow-based detection methods covering diverse detection paradigms:
(1) NetBeacon~\cite{netbeacon-zhou2023efficient}, a supervised tree-based per-flow detector;
(2) FSC~\cite{whisper}, a K-Means-based per-flow anomaly detector;
(3) Trident~\cite{zhao2024trident}, an autoencoder-based per-flow anomaly detector;
(4) NetVigil~\cite{hsieh2024netvigil}, a GNN-based autoencoder to detect anomalous hosts;
(5) Hypervision~\cite{hypervision}, a flow interaction graph-based outlier detector; and
(6) BS-CNN, a sequence-based method using the same 1D-CNN encoder as Tanchun to classify behavior sequences without behavioral attribute supervision.

\textbf{Datasets.}
We use three widely adopted public datasets: \textit{CIC-IDS2017}, \textit{CIC-IDS2018}~\cite{cic18-sharafaldin2018toward}, and \textit{CIC-IoT2023}~\cite{neto2023ciciot2023}. These datasets cover diverse HTTP(S) DDoS attacks, including HTTP floods generated by tools such as HOIC, LOIC, and GoldenEye, as well as slow attacks such as Slowloris.

We further synthesize two complex attack scenarios that frequently occur in real-world environments and challenge existing detection systems.
(1) \textit{Encrypted Low-Rate HTTPS Flood (LRHF)}, where each bot generates only two requests per second to simulate the low-rate flooding observed in a real attack reported by Baidu in 2023~\cite{huawei_report}. We combine these attack flows with benign HTTP(S) traffic from the January 2024 MAWI trace~\cite{mawi_kenjiro2000traffic}.
(2) \textit{Multi-vector Attack}, which combines HOIC, LOIC, GoldenEye, and Slowloris from CIC-IDS2018 in equal proportions, reflecting practical DDoS campaigns that employ multiple attack vectors simultaneously~\cite{huawei_report}.

Based on these datasets, we further construct controlled scenarios to evaluate different forms of traffic drift, as described in the corresponding evaluation subsections.

Finally, we validate Tanchun using real HTTP(S) DDoS attack traces and production HTTP(S) traffic collected from an operational web server. Details are presented in \S~\ref{real_trace}.

\textbf{Evaluation Procedure and Metrics}.
Following our detection system, all methods detect HTTP(S) flows using only the flow statistics described in \S~\ref{subsec_flow_system}. To prevent temporal data leakage, each dataset is chronologically split into training and test sets with an 8:2 ratio, where earlier flows are used for training and later flows for testing. We evaluate flow-level detection performance using accuracy and F1-score.

\subsection{Overall Performance}

We report the overall comparison between Tanchun and baseline methods across public datasets, synthetic complex attacks, and traffic drift scenarios. We summarize two representative drifts: benign traffic drift (\S~\ref{subsec_Robustness Under Benign Traffic Drifts}), which evaluates methods on background traffic collected one week after training, and cross-attack event (\S~\ref{subsec_Attack Pattern Drifts}), which evaluates generalization from known to unseen attacks. Detailed settings and results are provided in the following subsections.

As shown in Table~\ref{tab:overall}, \textbf{Tanchun achieves the best performance} across all settings, with an overall F1-score of 0.988. Notably, \textbf{Tanchun shows strong robustness to traffic drift}, outperforming the second-best method by 5.3\% and 9.6\% under benign and attack drift scenarios, respectively.

Tanchun achieves F1-scores above 0.99 on public datasets and remains effective under challenging synthetic attacks. In particular, under the LRHF, where attackers reduce request rates to mimic benign users, Tanchun achieves an F1-score of 0.999, demonstrating its ability to capture discriminative access behaviors beyond traffic volume. Other baselines show different weaknesses: anomaly detection methods (e.g., FSC, Trident, NetVigil, and HyperVision) struggle with attacks resembling benign traffic, especially under LRHF, while NetBeacon is limited by single-flow feature discrimination.

Tanchun achieves F1-scores of 0.982 and 0.955 under benign and attack drift scenarios, respectively. Other baselines are more sensitive to traffic drift: anomaly detection methods struggle with benign traffic evolution, while supervised methods fail to generalize to unseen attack features. Although BS-CNN captures temporal patterns through behavior sequences, it still underperforms Tanchun by 8.9\% and 8.8\% under benign and attack drift scenarios, respectively. These results demonstrate the benefit of behavioral attribute supervision.

\subsection{Robustness Under Benign Traffic Drift}
\label{subsec_Robustness Under Benign Traffic Drifts}
\begin{figure}
    \centering
    \includegraphics[width=0.8\linewidth]{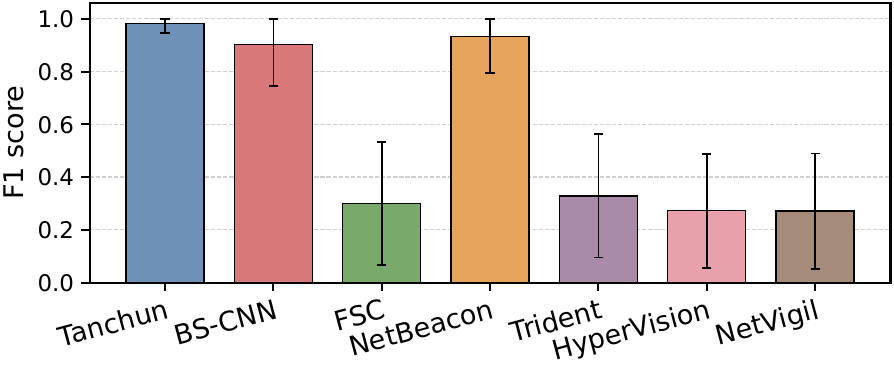}
    \caption{F1-score under benign traffic drift.}
    \label{fig:bg_drift}
\end{figure}

We use MAWI~\cite{mawi_kenjiro2000traffic} traces to simulate benign traffic evolution, as they provide real-world benign traffic collected from the same network node on different days. Specifically, we use traces from 2024.1.1 and 2024.1.8, separated by one week, as background traffic. Each trace is combined with representative HTTP(S) DDoS attacks (HOIC, LOIC, GoldenEye, and Slowloris) from CIC datasets~\cite{cic18-sharafaldin2018toward} to construct datasets. Tanchun is trained on the 2024.1.1 dataset and evaluated on the 2024.1.8 dataset. The results are shown in Fig.~\ref{fig:bg_drift}.

Tanchun demonstrates strong robustness against benign traffic drift, achieving an F1-score close to 1.0 with only minor degradation. This robustness stems from its ability to capture stable IP-pair access behaviors: although traffic statistics may evolve over time, behavioral differences between benign users and attackers remain distinguishable. In contrast, BS-CNN, which only models raw behavior sequences, suffers an 8.2\% F1-score drop compared to Tanchun, highlighting the benefit of behavioral attribute supervision. NetBeacon remains effective as its learned attack-flow boundaries largely persist under benign traffic drift. However, shifted benign flows may cross these boundaries, causing false positives and leading to lower performance than Tanchun. Anomaly detection methods (e.g., FSC, Trident, NetVigil, and HyperVision) are more sensitive to benign traffic variations, as shifted benign distributions may be incorrectly identified as anomalies.

% \begin{figure*}[t]
%     \centering

%     \begin{subfigure}[b]{0.32\textwidth}
%         \centering
%         \includegraphics[width=\linewidth]{chart/security_0101_to_0108_f1_mean_std_bar.pdf}
%         \caption{F1-score under Benign Traffic Drift.}
%         \label{fig:bg_drift}
%     \end{subfigure}
%     \hfill
%     \begin{subfigure}[b]{0.32\textwidth}
%         \centering
%         \includegraphics[width=\linewidth]{chart/security_rate_drift_LOIC.pdf}
%         \caption{F1-score with reduced bot attack rate.}
%         \label{fig:attack_rate_drift}
%     \end{subfigure}
%     \hfill
%     \begin{subfigure}[b]{0.32\textwidth}
%         \centering
%         \includegraphics[width=\linewidth]{chart/security_flow_statistics_drift.pdf}
%         \caption{F1-score with varying attack flow statistics.}
%         \label{fig:statistics_drift}
%     \end{subfigure}

%     \caption{F1-score under different drift settings.}
%     \label{fig:drift_comparison}
% \end{figure*}

\subsection{Robustness Under Attack Pattern Drift}
\label{subsec_Attack Pattern Drifts}

% \begin{figure*}[t]
%     \centering

%     \begin{minipage}[b]{0.3\textwidth}
%         \centering
%         \includegraphics[width=\linewidth]{chart_attack_drift/security_rate_drift_LOIC.pdf}
%         \caption{F1-score under attack-rate reduction.}
%         \label{fig:attack_rate_drift}
%     \end{minipage}
%     \hfill
%     \begin{minipage}[b]{0.3\textwidth}
%         \centering
%         \includegraphics[width=\linewidth]{chart_attack_drift/security_temporal_drift.pdf}
%         \caption{F1-score under random attack pauses.}
%         \label{fig:attack pause}
%     \end{minipage}
%         \hfill
%     \begin{minipage}[b]{0.36\textwidth}
%         \centering
%         \includegraphics[width=\linewidth]{chart_attack_drift/security_Tanchun_BS-CNN_zeroshot_heatmap.pdf}
%         \caption{Cross-event F1-score transfer.}
%         \label{fig:zeroshot}
%     \end{minipage}

% \end{figure*}

\begin{figure*}[t]
    \centering

    \begin{subfigure}[b]{0.30\textwidth}
        \centering
        \includegraphics[width=\linewidth]{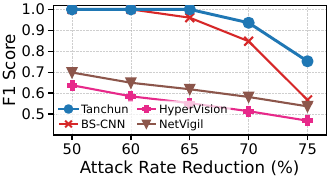}
        \caption{F1-score under attack-rate reduction.}
        \label{fig:attack_rate_drift}
    \end{subfigure}
    \hfill
    \begin{subfigure}[b]{0.30\textwidth}
        \centering
        \includegraphics[width=\linewidth]{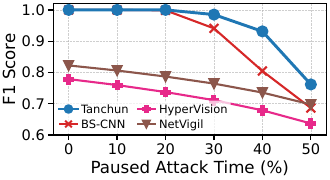}
        \caption{F1-score under random attack pauses.}
        \label{fig:attack_pause}
    \end{subfigure}
    \hfill
    \begin{subfigure}[b]{0.36\textwidth}
        \centering
        \includegraphics[width=\linewidth]{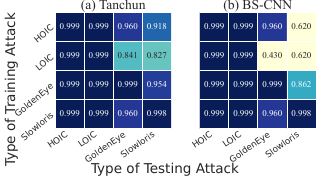}
        \caption{F1-score under cross-event attack.}
        \label{fig:zeroshot}
    \end{subfigure}

    \caption{Robustness under different attack drift scenarios.}
    \label{fig:attack_drift}
\end{figure*}

\textbf{Attack Rate Reduction.} We simulate a common evasion strategy where attackers reduce the request rate of individual bots to make attack traffic resemble benign access~\cite{huawei_report}. We use the LOIC dataset, an HTTP flood trace, for evaluation. We train the model on the original dataset and randomly reduce the number of attack flows generated by each bot in the test set to simulate different request rate reductions. Since per-flow detection methods are independent of such aggregated traffic changes, we exclude them from this evaluation. The F1-score variation under different attack rates is shown in Fig.~\ref{fig:attack_rate_drift}.

Tanchun achieves the best F1-score across different levels of attack-rate reduction, with BS-CNN ranking second. This suggests that, despite the reduced attack rate, the temporal pattern differences between bots and benign users remain distinguishable. Under severe rate reduction (i.e., 75\%), Tanchun outperforms BS-CNN by 32.7\%. This advantage stems from attribute alignment, which helps Tanchun capture rate-independent behavioral characteristics, such as regular request patterns, and learn more robust representations. In contrast, HyperVision and NetVigil are more sensitive to rate reduction because their detection relies heavily on traffic-volume-related features, such as the aggregated IP-pair flow counts, which are affected by the decrease in attack rate.

% \begin{figure}
%     \centering
%     \includegraphics[width=0.8\linewidth]{chart/security_rate_drift_LOIC.pdf}
%     \caption{Variation of F1-score with reducing bot attack rate.}
%     \label{fig:attack rate drift}
% \end{figure}

\textbf{Intermittent Attacks.}
Attackers can evade temporal behavior analysis by intermittently pausing their attacks, e.g., through pulse attacks~\cite{accturbo-alcoz2022aggregate}. Such pauses reduce access activity and temporal regularity, making bots resemble benign users and challenging Tanchun’s relevant attributes. To simulate this behavior, we randomly insert one-second pauses into LOIC attack traces while keeping benign flows unchanged, varying the paused fraction from 0\% to 50\%. All models are trained on the original traces and evaluated under different pause ratios. Fig.~\ref{fig:attack_pause} shows the F1-score changes of the four methods that consider IP-pair behavior.

It can be seen that Tanchun maintains the best F1-score as the attack-pause ratio increases. BS-CNN achieves the second-best performance, but its F1-score is 13.6\% lower than Tanchun at a 40\% pause ratio. This is because frequent pauses substantially distort the raw behavior sequences learned by BS-CNN, leading to classification errors. In contrast, Tanchun additionally leverages behavioral attributes. Although \textit{Activity Persistence}, \textit{Activity Continuity}, and \textit{Temporal Regularity} are affected by pauses, more stable semantics such as \textit{Flow Consistency} still help distinguish attackers from benign users. HyperVision and NetVigil perform substantially worse than Tanchun, mainly because more frequent pauses reduce observable attack rate, making malicious traffic more similar to benign behavior and resulting in more false negatives.

\textbf{Cross-Attack Events.}
Attackers may employ different tools or strategies across attack events, leading to variations in attack behaviors over time. To evaluate this scenario, we use four HTTP(S) DDoS events from IDS18~\cite{cic18-sharafaldin2018toward}, including LOIC, HOIC, GoldenEye, and Slowloris, generated by different attack tools with benign background traffic collected from the same network environment. Each method is trained on one attack event and tested on the others. The cross-attack event performance is summarized in Table~\ref{tab:overall}. In addition, Fig.~\ref{fig:zeroshot} presents the cross-event transfer matrices of Tanchun and the best-performing baseline, while the complete matrices of other methods are provided in Appendix~\ref{whole_zeroshot}.

Tanchun consistently achieves the best performance across all training-testing combinations, demonstrating strong robustness to unseen attack vectors by capturing stable malicious behaviors shared across attacks. In contrast, BS-CNN yields an average F1-score 8.8\% lower and fails in three combinations, highlighting the value of behavioral attribute supervision for learning robust semantics. NetBeacon nearly loses its detection capability when attack-flow features differ substantially from those seen during training. Trident, FSC, and HyperVision struggle with Slowloris because its flow characteristics closely resemble benign traffic. NetVigil performs well in most cases through IP-pair aggregation and graph modeling, but remains less effective on Slowloris, indicating that its graph-based modeling is less effective at capturing the temporal dynamics of Slowloris attacks.

\subsection{Ablation Study}
\label{sec_ablation}

In this subsection, we evaluate the effectiveness of Tanchun’s core components through ablation studies. We consider benign traffic drift and cross-attack event scenarios, where the latter serves as a representative case of attack drift.

% \begin{figure}
%     \centering
%     \begin{subfigure}[b]{0.48\linewidth}
%         \includegraphics[width=\textwidth]{chart/ablation_loic.pdf}
%         \caption{Train on LOIC.}
%         \label{fig:ablation_loic}
%     \end{subfigure}
%     \hfill
%     \begin{subfigure}[b]{0.48\linewidth}
%         \includegraphics[width=\textwidth]{chart/ablation_goldeneye.pdf}
%         \caption{Train on GoldenEye.}
%         \label{fig:ablation_goldeneye}
%     \end{subfigure}

%     \caption{Ablation Study on Contrastive Learning Loss and Behavioral Descriptions.}
%     \label{fig:ablation}
% \end{figure}

\begin{table}[t]
    \centering
    \caption{Ablation study on the key modules.}
    \label{tab:ablation}

    \setlength{\tabcolsep}{4.0pt}
    \renewcommand{\arraystretch}{0.8}

    \resizebox{\columnwidth}{!}{%
    \begin{tabular}{llcc}
        \toprule
        \textbf{Study}
        & \textbf{Variant}
        & \textbf{Benign Drift}
        & \textbf{Attack Drift} \\
        \midrule

        BSA (\S~\ref{subsec_sequence_Abstraction})
        & w/o BS
        & 0.980
        & \textcolor{red}{0.292} \\

        \midrule

        ARL (\S~\ref{subsec_ARL})
        & w/o $L_{\mathrm{align}}$
        & 0.902
        & 0.871 \\

        \midrule

        \multirow{4}{*}{\makecell{Attribute\\Utilization}}
        & Attribute-only
        & 0.819
        & 0.721 \\

        & Indicator-only
        & 0.764
        & 0.896 \\

        & Indicator Concat.
        & 0.909
        & 0.851 \\

        & Attr. Pred.
        & 0.982
        & 0.871 \\

        \midrule

        Full
        & \textbf{Tanchun}
        & \textbf{0.982}
        & \textbf{0.955} \\

        \bottomrule
    \end{tabular}%
    }
\end{table}

\textbf{Study on Behavior Sequence Abstraction.}
To validate the effectiveness of behavior sequences for temporal analysis, we construct a variant, \textbf{w/o BS}, that removes the time-slot-based sequence construction. Specifically, we aggregate the features of each IP pair over the entire observation window of length $L$ and replace the 1D-CNN with an MLP of the same depth, since sequential modeling is no longer required. The results are shown in Table~\ref{tab:ablation}. Removing behavior sequence substantially degrades performance under attack drift, with the F1-score dropping from 0.955 to 0.292. This result shows that, compared with coarse-grained IP-pair aggregation, behavior sequences better preserve temporal access patterns that are important for identifying diverse attack behaviors.

\textbf{Study on Attribute-guided Representation Learning.}
We evaluate whether the proposed attribute alignment task ($L_{\mathrm{align}}$) improves the generalizability of Tanchun. We set the weight $\lambda$ of $L_{\mathrm{align}}$ to 0 while keeping the remaining model unchanged, denoted as \textbf{w/o $L_{\mathrm{align}}$}. As shown in Table~\ref{tab:ablation}, the F1-score decreases from 0.982 to 0.902 under benign traffic drift and from 0.955 to 0.871 under attack drift. These results indicate that the behavioral abstractions provided by the attributes can serve as effective supervision signals, guiding the model toward more generalizable representations.

\textbf{Alternative Utilization of Behavioral Attributes.}
We compare four alternative ways of utilizing behavioral attributes.
\textbf{(1) Attribute-only:} We directly detect attacks using the multi-hot attribute vector generated with the same criteria as in training.
\textbf{(2) Indicator-only:} We use the seven quantitative indicators corresponding to the behavioral attributes as detection features.
\textbf{(3) Indicator Concat.:} We concatenate the seven indicators with the behavior sequence representation for detection.
\textbf{(4) Attr. Pred.:} We replace attribute alignment with a conventional auxiliary task that predicts behavioral attributes from the behavior sequence representation.
As shown in Table~\ref{tab:ablation}, none of these variants consistently matches Tanchun across both drift scenarios. The Attribute-only and Indicator-only variants fail to retain fine-grained temporal patterns, while concatenation and attribute prediction provide only limited semantic supervision. In contrast, Tanchun's attribute alignment directly constrains the representation space, enabling more generalizable behavioral semantics.

\subsection{Parameter Analysis}

\begin{figure*}[t]
    \centering

    \begin{subfigure}[t]{0.32\textwidth}
        \centering
        \includegraphics[width=\linewidth]{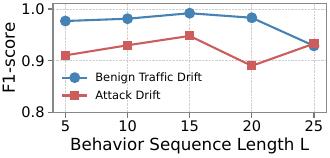}
        \caption{Analysis of behavior sequence length.}
        \label{fig:eva_length_analysis}
    \end{subfigure}
    \hfill
    \begin{subfigure}[t]{0.32\textwidth}
        \centering
        \includegraphics[width=\linewidth]{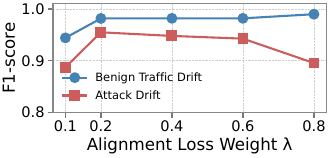}
        \caption{Analysis of alignment loss weight.}
        \label{fig:analysis weight}
    \end{subfigure}
    \hfill
    \begin{subfigure}[t]{0.32\textwidth}
        \centering
        \includegraphics[width=\linewidth]{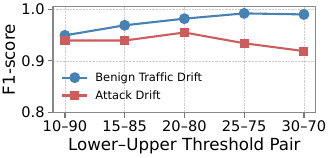}
        \caption{Analysis of level assignment threshold.}
        \label{fig:threshold}
    \end{subfigure}

    \caption{Sensitivity analysis of three key parameters.}
    \label{fig:parameters}
\end{figure*}

In this subsection, we use the same scenarios as in \S~\ref{sec_ablation} to analyze the impact of several key parameters.

\textbf{Analysis on Behavior Sequence Length.} 
\footnote{We keep $\Delta t$ fixed, as $\Delta t$ and $L$ jointly determine the temporal observation horizon. We therefore analyze $L$ to avoid redundant parameter combinations.}
We vary the length $L$ of the behavior sequence from 5 to 25, and the changes in Tanchun’s F1-score are shown in Fig.~\ref{fig:eva_length_analysis}. As observed, setting L=15 achieves the best F1-score in both scenarios. Under the benign traffic drift and attack drift scenarios, it improves the performance by 6.8\% and 6.5\%, respectively, compared with the worst cases. This is because when $L$ is set too small, the available temporal information is insufficient to distinguish malicious behaviors. In contrast, when $L$ is set too large, excessive redundant information is introduced, which may hinder the model from capturing essential patterns. Therefore, we set $L=15$ in our implementation.

\textbf{Analysis of the Attribute Alignment Loss Weight.}
We vary the weight $\lambda$ of $L_{\text{align}}$ from 0.1 to 0.8, with the resulting F1-scores shown in Fig.~\ref{fig:analysis weight}. Under benign traffic drift, the F1-score improves as $\lambda$ increases to 0.2 and then remains nearly stable. Under attack drift, the F1-score also reaches its highest level at $\lambda=0.2$, remains on a plateau, and drops sharply at $\lambda=0.8$. The plateau improves F1-score by 7.6\% and 6.6\% over $\lambda=0.1$ and $\lambda=0.8$, respectively. This indicates that a moderate $\lambda$ promotes more generalizable behavioral semantics, while an overly large weight weakens fine-grained temporal modeling. We therefore set $\lambda=0.2$ as the default.

\textbf{Analysis of Level Assignment Threshold.}
We vary the threshold pair $(\theta_a^{\mathrm{low}}, \theta_a^{\mathrm{high}})$ from 10--90 to 30--70 while keeping other settings fixed. For the unbounded attribute \textit{Access Intensity}, these values denote percentiles, while for bounded indicators, 10--90 corresponds to normalized thresholds of 0.1 and 0.9, respectively.  As shown in Fig.~\ref{fig:threshold}, under benign traffic drift, the F1-score increases from 0.949 at 10--90 to 0.992 at 25--75 and then remains nearly stable. Under attack drift, performance peaks at 20--80 with an F1-score of 0.955, but drops to 0.919 at 30--70. A larger gap between $\theta_a^{\mathrm{low}}$ and $\theta_a^{\mathrm{high}}$ assigns more samples to the medium level, weakening supervision from the low and high levels, whereas a smaller gap may separate similar borderline behaviors into different levels and introduce assignment ambiguity. We therefore use 20--80 as the default setting, which provides the best overall balance across the two drift scenarios.

% \begin{figure}
%     \centering
%     \includegraphics[width=0.7\linewidth]{chart/security_ablation_seq_len_f1.pdf}
%     \caption{Analysis on Behavior Sequence Length.}
%     \label{fig:eva_length_analysis}
% \end{figure}

\subsection{Representation Analysis}
\label{subsec_representation_analysis}

\begin{figure}
    \centering
    \begin{subfigure}[b]{0.48\linewidth}
        \includegraphics[width=\textwidth]{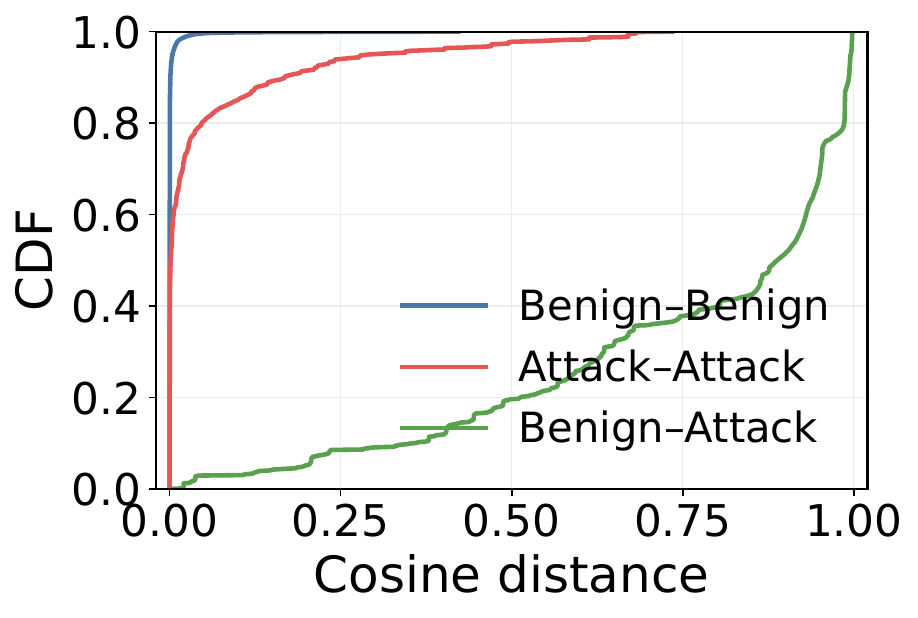}
        \caption{Distance Distribution}
        \label{fig:distance_cdf}
    \end{subfigure}
    \hfill
    \begin{subfigure}[b]{0.48\linewidth}
        \includegraphics[width=\textwidth]{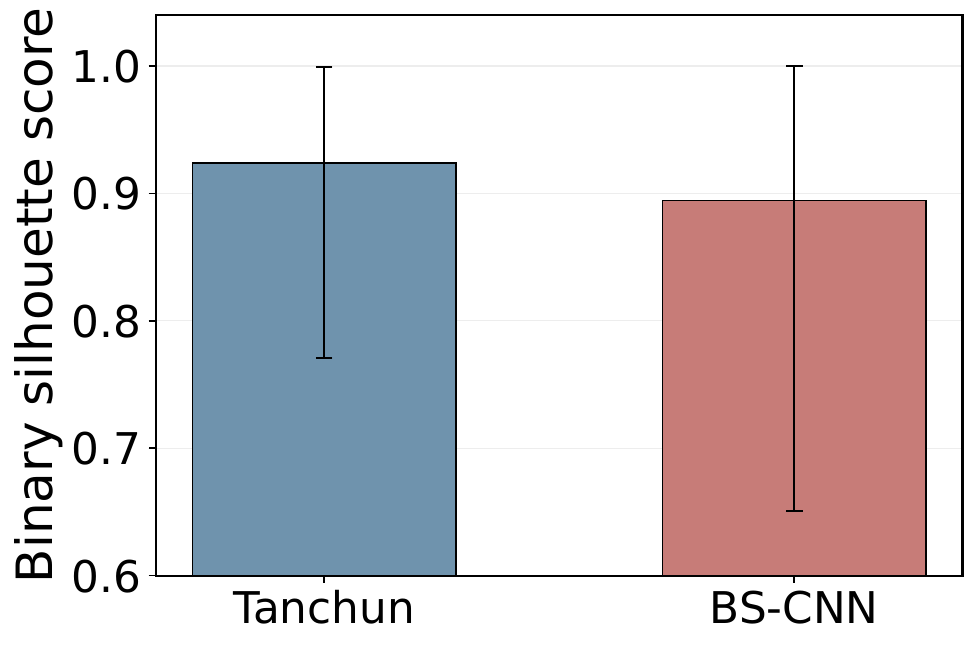}
        \caption{Binary Silhouette Score}
        \label{fig:Binary_Silhouette}
    \end{subfigure}

    \caption{Analysis of the behavior sequence representation.}
    \label{fig:rep_distence}
\end{figure}

We analyze Tanchun's learned behavior sequence representations $h$ under 12 cross-attack event training-testing combinations. For each combination, we compute cosine distances among benign--benign, attack--attack, and benign--attack representation pairs. As shown in Fig.~\ref{fig:distance_cdf}, benign and attack representations have small intra-class distances but much larger inter-class distances, indicating that Tanchun learns compact and well-separated representations. Such representation structure supports robust detection under traffic drift.

We further quantify the clustering quality of the learned representations using the binary silhouette score~\cite{rousseeuw1987silhouettes}, which jointly measures intra-class compactness and inter-class separation. Specifically, for each of the 12 cross-event training-testing combinations, we compute the silhouette score between benign and attack representations and compare Tanchun with BS-CNN. As shown in Fig.~\ref{fig:Binary_Silhouette}, Tanchun improves the average silhouette score from $0.894$ to $0.924$ and the worst-case score from $0.651$ to $0.771$, demonstrating that behavioral attribute supervision yields more compact and separable representations under traffic drift.

\subsection{Evaluation with Real-World Traffic}
\label{real_trace}

We evaluate Tanchun in a deployment-oriented testbed using real-world HTTP(S) production traffic and DDoS attack traces. The benign traffic is collected from an operational campus web server cluster hosting multiple websites and contains 317,010 HTTP(S) flows. The attack traffic comprises five real HTTP(S) DDoS traces, including two HTTPS floods and three HTTP floods, collected from a top-tier DDoS scrubber~\cite{huawei_12000} and verified by security operators. In the testbed, we continuously replay production traffic and inject each attack trace separately, allowing the system to process mixed benign and attack flows in a deployment-relevant setting.

\begin{figure}
    \centering
    \begin{subfigure}[b]{0.49\linewidth}
        \includegraphics[width=\textwidth]{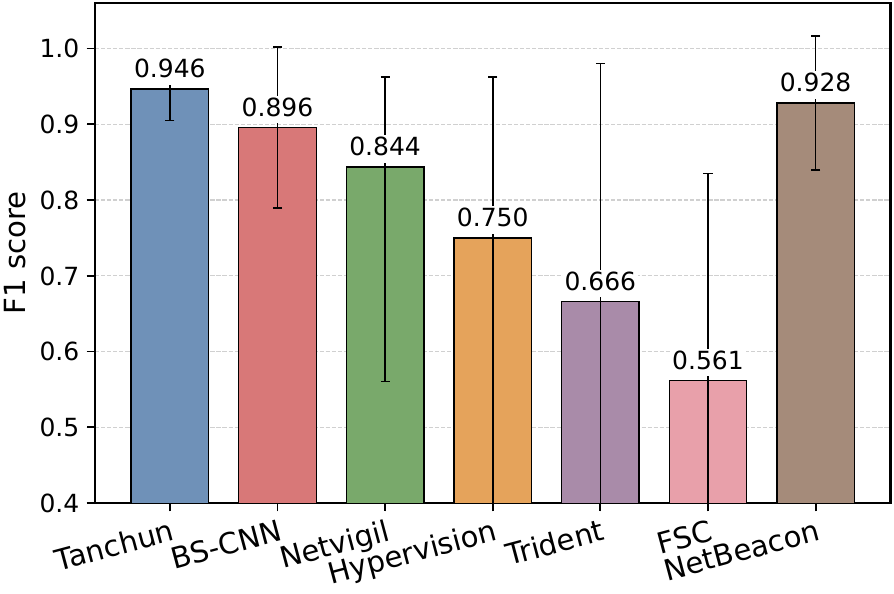}
        \caption{Comparison Evaluation}
        \label{fig:real_trace_bar}
    \end{subfigure}
    \hfill
    \begin{subfigure}[b]{0.49\linewidth}
        \includegraphics[width=\textwidth]{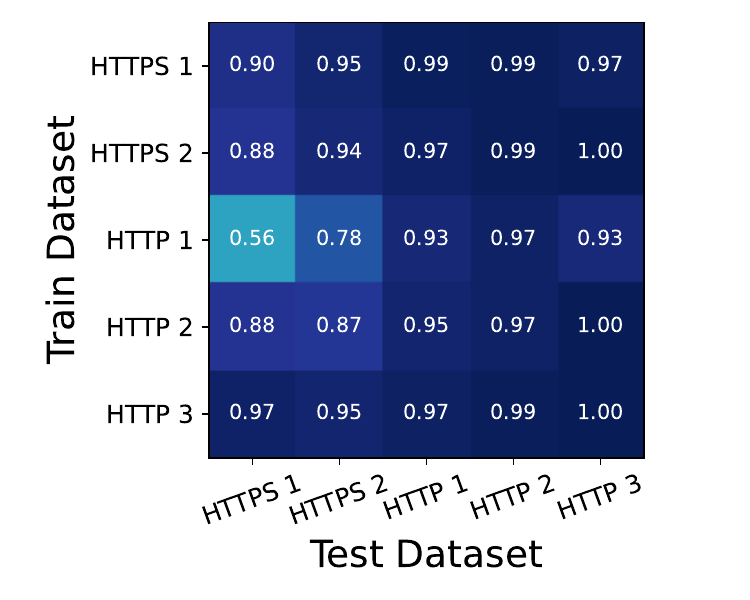}
        \caption{Cross-Attack Event}
        \label{fig:real_trace_zeroshot}
    \end{subfigure}

    \caption{Detection results on real-world traffic trace.}
    \label{fig:real_world_results}
\end{figure}

We conduct two experiments: (i) known-attack detection, where Tanchun is compared with existing baselines; and (ii) unseen-attack detection, where Tanchun is trained on one attack type and evaluated on the remaining unseen attack types.
As shown in Figure~\ref{fig:real_trace_bar}, Tanchun achieves the highest average F1-score across all five attacks, outperforming the baselines by 22.3\% on average. This result suggests that Tanchun remains effective on real-world attack traces. As shown in Figure~\ref{fig:real_trace_zeroshot}, Tanchun also maintains F1-scores close to or above 0.90 for most train-test pairs, indicating promising generalization to unseen attacks. Overall, these results suggest that Tanchun can maintain reliable detection performance on real-world HTTP(S) DDoS.

\subsection{Efficiency Evaluation} 
\label{sec:efficiency}
\begin{figure}
    \centering
    \begin{subfigure}[b]{0.48\linewidth}
        \includegraphics[width=\textwidth]{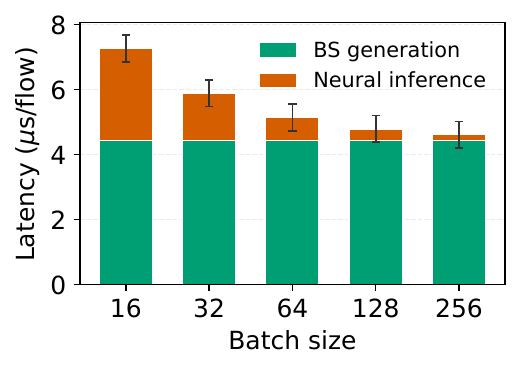}
        \caption{Latency of Each Stage.}
        \label{fig:tanchun_latency}
    \end{subfigure}
    \hfill
    \begin{subfigure}[b]{0.48\linewidth}
        \includegraphics[width=\textwidth]{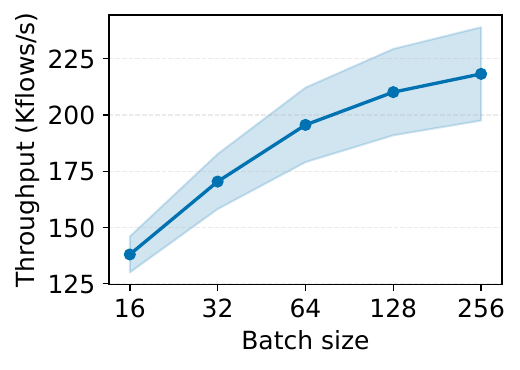}
        \caption{Detection Throughput.}
        \label{fig:tanchun_tp}
    \end{subfigure}

    \caption{Tanchun efficiency evaluation.}
    \label{fig:efficiency}
\end{figure}

When the flows to be detected enter the system,
Tanchun’s detection consists of two main stages: behavior sequence generation on the CPU and neural network inference on the GPU. We separately measure the latency of these two stages and evaluate the overall throughput. Since neural inference can benefit from batch processing, we vary the batch size from 16 to 256, with the results shown in Fig.~\ref{fig:efficiency}.

\textbf{Latency.}
As shown in Fig.~\ref{fig:tanchun_latency}, behavior sequence generation introduces a fixed overhead of approximately $4.43 \mu$s per flow. Neural network inference benefits from batch processing, and its amortized latency decreases significantly as the batch size increases. Specifically, increasing the batch size from 16 to 256 reduces the neural inference latency from $2.8 \mu$s to only $0.18 \mu$s per flow. At a batch size of 256, Tanchun achieves an overall detection latency of $4.61 \mu$s per flow.

\textbf{Throughput.}
As shown in Fig.~\ref{fig:tanchun_tp}, Tanchun’s detection throughput increases with the batch size and reaches 218K flows/s at a batch size of 256. However, further increasing the batch size provides limited improvement, as behavior sequence generation becomes the bottleneck. Given that Internet services may generate around 100K flows per second~\cite{lititanlog}, these results show that Tanchun can process such large-scale flows in real time.

\section{Discussion}

\textbf{Training Data for Tanchun.}
The training data is generally available in practical deployments.
Benign flows can be directly collected from the target network through existing flow telemetry systems~\cite{netflow_claise2004cisco,Securing_Public_Cloud_mani2025securing}.
For attack flows, existing HTTP(S) defense mechanisms, such as decryption-based detectors~\cite{huawei_report} and JS challenge systems~\cite{JS-challenge}, can provide high-confidence malicious samples.
By leveraging these existing detection results during training, Tanchun can learn attack behaviors while reducing the reliance on costly decryption and minimizing disruptions to benign users caused by JS challenges.

\textbf{Consideration on Behavioral Attributes.} Tanchun defines behavioral attributes based on domain knowledge of common attack and benign behaviors. These attributes serve as stable semantic anchors to learn robust behavioral semantics under traffic drift. As supervision signals rather than detection rules, they allow Tanchun to tolerate attribute-level annotation noise caused by threshold-based assignment uncertainty (Appendix~\ref{app_noise}). Attribute quality and coverage can be further improved through operational feedback and automated attribute discovery techniques~\cite{patel2025agua}.

% \textbf{Behavior Attribution.} Existing approaches have also leveraged large language models (LLMs) to automate the attribution of network data. However, in our exploration, we found that the inherent randomness of LLM outputs results in inconsistent labels, undermining the reliability of the attribution. In this paper, Tanchun adopts an indicator-based attribution approach that enables objective and quantitative attribution. Moreover, human verification on sampled results can further enhance the accuracy and reliability of the attribution process.

% \textbf{Detect Other Attacks.} Tanchun has the potential to be extended to other application-layer attacks which similarly aim to overwhelm or invade servers with excessive requests, such as DNS flooding~\cite{DNS_flood_alonso2016mining} and brute-force~\cite{brute-force_hofstede2017flow}.

\textbf{Collective Behavior Analysis.} 
In HTTP(S) DDoS attacks, bots from the same campaign may exhibit similar behavioral patterns, providing additional signals for identifying coordinated attacks. Tanchun learns a discriminative representation space where malicious IP-pairs with similar access behaviors tend to cluster together (\S~\ref{subsec_representation_analysis}). Future work can leverage cluster-level characteristics, such as cluster density and behavioral consistency, to further enhance detection.

\textbf{Co-Optimization with Flow Telemetry.} Further improvements in end-to-end detection efficiency can be achieved through flow telemetry optimization. Reducing the flow export interval to the second level~\cite{netflow_export} can better align telemetry granularity with Tanchun's behavior sequence construction. Moreover, adaptive configuration of flow active and inactive timeouts~\cite{netflow_export} can balance detection timeliness and flow completeness. Shorter inactive timeouts can reduce delays caused by long-lived connections, such as Slowloris, while shorter active timeouts can enable earlier detection, since early-stage flows may already provide sufficient information for IP-pair analysis. We will explore this trade-off in future work.

\textbf{Detect Other Attacks.} The design principle of Tanchun may apply to other attacks that exhibit malicious access behavioral patterns, such as DNS query flooding~\cite{DNS_flood_alonso2016mining} and brute-force attacks~\cite{brute-force_hofstede2017flow}. Future work will explore attack-specific behavioral attributes and evaluate their effectiveness across broader attack scenarios.
\section{Related Work}
\textbf{HTTP(S) DDoS Defense}. Liao et al.~\cite{http_related_seq_liao2015application} detect malicious users by modeling inter-request patterns. Xie et al.~\cite{http_related_websit_xie2008monitoring,http_related_markov_xie2008large} analyze website access behaviors and request relationships using hidden semi-Markov models. Singh et al.~\cite{http_related_user_behavior_singh2018user} further characterize user behaviors based on request frequency and resource repetition. However, these approaches rely on application-layer information, which typically requires HTTPS decryption and introduces additional decryption overhead and may raise privacy concerns.

\textbf{Volumetric DDoS Defense}. Some methods detect DDoS using flow data collected from IXP or ISPs. IXP-Scrubber~\cite{wichtlhuber2022ixp} learns flow features from blackholing traffic, and Xatu~\cite{xu2022xatu} leverages auxiliary signals to enhance detection.
Some methods defend against DDoS on programmable switches. Poseidon~\cite{zhang2020poseidon} and Jaqen~\cite{liu2021jaqen} mitigate volumetric DDoS attacks by deploying rules. Ripple~\cite{link_xing2021ripple} and Mew~\cite{DDoS_related_mew} mitigate link-flooding attacks~\cite{link_coremelt,link_kang2013crossfire} by analyzing multi-switch traffic. ACC-Turbo~\cite{accturbo-alcoz2022aggregate} defends against pulse-wave DDoS by aggregate-based congestion control. However, HTTP(S) DDoS do not usually consume excessive bandwidth, making it difficult to trigger the volumetric DDoS defenses.

\textbf{Encrypted Traffic Detection}. Kitsune~\cite{mirsky2018kitsune} and nPrintML~\cite{nprintML} perform packet-level detection using AutoEncoder and AutoML, respectively. NetBeacon~\cite{netbeacon-zhou2023efficient} and FlowLens~\cite{barradas2021flowlens} use tree models to classify traffic based on flow statistics, Trident~\cite{zhao2024trident} learns flow representations under class-incremental settings, and Whisper~\cite{whisper} captures frequency-domain characteristics of flows. However, HTTP(S) DDoS attacks often consist of seemingly legitimate requests, limiting the discriminative power of packet- and flow-level features. HyperVision~\cite{hypervision} and NetVigil~\cite{hsieh2024netvigil} further model inter-flow relationships and an IP-pair–level graph, respectively. Nevertheless, neither explicitly captures attackers' temporal access patterns, potentially overlooking malicious temporal behaviors.

% \textbf{Image Detection via Text Semantics.} OOD and Anomaly detection~\cite{ood_detection_survey_xu2024large} are typically studied in the field of Computer Vision: the former detects anomalies within an image, while the latter determines whether an image lies outside the training data distribution. This is analogous to our task of malicious traffic detection. ALFA~\cite{related_ALFA_zhu2024llms} generates descriptions of normal and anomalous objects using LLMs and detects the presence of anomalies by comparing this textual information with the image. Dai et al.~\cite{related_ood_dai2023exploring} prompt LLMs to obtain distinctive characteristics for each object category and use them to determine whether certain images are OOD. TagFog~\cite{related_chen2024tagfog} generates descriptions for different image classes using LLMs and aligns image representations with these descriptions to enhance the detection of OOD samples. However, how to leverage textual information to enhance traffic analysis for malicious traffic detection remains an open problem. To our knowledge, we are the first to explore this direction.

\section{Conclusion}
This paper presents Tanchun, a novel flow-based HTTP(S) DDoS detection framework that leverages human-understandable behavioral attributes as semantic supervision. Tanchun first models the temporal access behaviors of IP-pairs from flow statistics and derives behavioral attributes from multiple perspectives to comprehensively characterize access behavior. Through attribute-guided representation learning, Tanchun aligns traffic representations with behavioral semantics, thereby improving the robustness of HTTP(S) DDoS detection under traffic drift.

% \subfile{sections/ethical concern}
% \subfile{sections/acknowledgements}

%-------------------------------------------------------------------------------
% optional clearing of the page
\cleardoublepage

\cleardoublepage

\bibliographystyle{plainurl}
\bibliography{tanchun}

@inproceedings{zhang2020poseidon,
  title={Poseidon: Mitigating volumetric ddos attacks with programmable switches},
  author={Zhang, Menghao and Li, Guanyu and Wang, Shicheng and Liu, Chang and Chen, Ang and Hu, Hongxin and Gu, Guofei and Li, Qianqian and Xu, Mingwei and Wu, Jianping},
  booktitle={the 27th Network and Distributed System Security Symposium (NDSS 2020)},
  year={2020}
}

@inproceedings{liu2021jaqen,
  title={Jaqen: A $\{$High-Performance$\}$$\{$Switch-Native$\}$ approach for detecting and mitigating volumetric $\{$DDoS$\}$ attacks with programmable switches},
  author={Liu, Zaoxing and Namkung, Hun and Nikolaidis, Georgios and Lee, Jeongkeun and Kim, Changhoon and Jin, Xin and Braverman, Vladimir and Yu, Minlan and Sekar, Vyas},
  booktitle={30th USENIX Security Symposium (USENIX Security 21)},
  pages={3829--3846},
  year={2021}
}

@inproceedings{link_xing2021ripple,
  title={Ripple: A programmable, decentralized $\{$Link-Flooding$\}$ defense against adaptive adversaries},
  author={Xing, Jiarong and Wu, Wenqing and Chen, Ang},
  booktitle={30th USENIX Security Symposium (USENIX Security 21)},
  pages={3865--3881},
  year={2021}
}

@inproceedings{DDoS_related_mew,
  title={Mew: Enabling large-scale and dynamic link-flooding defenses on programmable switches},
  author={Zhou, Huancheng and Hong, Sungmin and Liu, Yangyang and Luo, Xiapu and Li, Weichao and Gu, Guofei},
  booktitle={2023 IEEE Symposium on Security and Privacy (SP)},
  pages={3178--3192},
  year={2023},
  organization={IEEE}
}

@inproceedings{link_kang2013crossfire,
  title={The crossfire attack},
  author={Kang, Min Suk and Lee, Soo Bum and Gligor, Virgil D},
  booktitle={2013 IEEE symposium on security and privacy},
  pages={127--141},
  year={2013},
  organization={IEEE}
}

@inproceedings{link_coremelt,
  title={The coremelt attack},
  author={Studer, Ahren and Perrig, Adrian},
  booktitle={European Symposium on Research in Computer Security},
  pages={37--52},
  year={2009},
  organization={Springer}
}

@inproceedings{accturbo-alcoz2022aggregate,
  title={Aggregate-based congestion control for pulse-wave DDoS defense},
  author={Alcoz, Albert Gran and Strohmeier, Martin and Lenders, Vincent and Vanbever, Laurent},
  booktitle={Proceedings of the ACM SIGCOMM 2022 Conference},
  pages={693--706},
  year={2022}
}

@inproceedings{wichtlhuber2022ixp,
  title={IXP scrubber: learning from blackholing traffic for ML-driven DDoS detection at scale},
  author={Wichtlhuber, Matthias and Strehle, Eric and Kopp, Daniel and Prepens, Lars and Stegmueller, Stefan and Rubina, Alina and Dietzel, Christoph and Hohlfeld, Oliver},
  booktitle={Proceedings of the ACM SIGCOMM 2022 Conference},
  pages={707--722},
  year={2022}
}

@inproceedings{mirsky2018kitsune,
  title={Kitsune: An Ensemble of Autoencoders for Online Network Intrusion Detection},
  author={Mirsky, Yisroel and Doitshman, Tomer and Elovici, Yuval and Shabtai, Asaf},
  booktitle={Network and Distributed Systems Security (NDSS) Symposium},
  year={2018}
}

@inproceedings{nprintML,
  title={New directions in automated traffic analysis},
  author={Holland, Jordan and Schmitt, Paul and Feamster, Nick and Mittal, Prateek},
  booktitle={Proceedings of the 2021 ACM SIGSAC Conference on Computer and Communications Security},
  pages={3366--3383},
  year={2021}
}

@inproceedings{barradas2021flowlens,
  title={FlowLens: Enabling Efficient Flow Classification for ML-based Network Security Applications.},
  author={Barradas, Diogo and Santos, Nuno and Rodrigues, Lu{\'\i}s and Signorello, Salvatore and Ramos, Fernando MV and Madeira, Andr{\'e}},
  booktitle={NDSS},
  year={2021}
}

@inproceedings{netbeacon-zhou2023efficient,
  title={An efficient design of intelligent network data plane},
  author={Zhou, Guangmeng and Liu, Zhuotao and Fu, Chuanpu and Li, Qi and Xu, Ke},
  booktitle={32nd USENIX Security Symposium (USENIX Security 23)},
  pages={6203--6220},
  year={2023}
}

@inproceedings{whisper,
  title={Realtime robust malicious traffic detection via frequency domain analysis},
  author={Fu, Chuanpu and Li, Qi and Shen, Meng and Xu, Ke},
  booktitle={Proceedings of the 2021 ACM SIGSAC Conference on Computer and Communications Security},
  pages={3431--3446},
  year={2021}
}

@article{hypervision,
  title={Detecting Unknown Encrypted Malicious Traffic in Real Time via Flow Interaction Graph Analysis},
  author={Fu, Chuanpu and Li, Qi and Xu, Ke}
}

@inproceedings{zhao2024trident,
  title={Trident: A Universal Framework for Fine-Grained and Class-Incremental Unknown Traffic Detection},
  author={Zhao, Ziming and Li, Zhaoxuan and Song, Zhuoxue and Li, Wenhao and Zhang, Fan},
  booktitle={Proceedings of the ACM on Web Conference 2024},
  pages={1608--1619},
  year={2024}
}

@inproceedings{http_related_rl_feng2020application,
  title={Application-layer DDoS defense with reinforcement learning},
  author={Feng, Yebo and Li, Jun and Nguyen, Thanh},
  booktitle={2020 IEEE/ACM 28th International Symposium on Quality of Service (IWQoS)},
  pages={1--10},
  year={2020},
  organization={IEEE}
}

@article{http_related_seq_liao2015application,
  title={Application layer DDoS attack detection using cluster with label based on sparse vector decomposition and rhythm matching},
  author={Liao, Qin and Li, Hong and Kang, Songlin and Liu, Chuchu},
  journal={Security and Communication Networks},
  volume={8},
  number={17},
  pages={3111--3120},
  year={2015},
  publisher={Wiley Online Library}
}

@article{http_related_user_behavior_singh2018user,
  title={User behavior analytics-based classification of application layer HTTP-GET flood attacks},
  author={Singh, Karanpreet and Singh, Paramvir and Kumar, Krishan},
  journal={Journal of Network and Computer Applications},
  volume={112},
  pages={97--114},
  year={2018},
  publisher={Elsevier}
}

@article{http_related_markov_xie2008large,
  title={A large-scale hidden semi-Markov model for anomaly detection on user browsing behaviors},
  author={Xie, Yi and Yu, Shun-Zheng},
  journal={IEEE/ACM transactions on networking},
  volume={17},
  number={1},
  pages={54--65},
  year={2008},
  publisher={IEEE}
}

@article{http_related_websit_xie2008monitoring,
  title={Monitoring the application-layer DDoS attacks for popular websites},
  author={Xie, Yi and Yu, Shun-Zheng},
  journal={IEEE/ACM Transactions on networking},
  volume={17},
  number={1},
  pages={15--25},
  year={2008},
  publisher={IEEE}
}

@misc{HTTP_multiplexing_flood,
  author = {Akamai},
  note = {\url{https://www.akamai.com/blog/security/akamai-protects-customers-http2-rapid-reset-ddos-attacks}},
  title = {
How Akamai Protects Customers from HTTP/2 Rapid Reset DDoS Attacks},
  year = {2025}
}

@techreport{netflow_claise2004cisco,
  title={Cisco systems netflow services export version 9},
  author={Claise, Benoit},
  year={2004}
}

@misc{huawei_report,
  author = {China Telecom Cybersecurity Technology Co. et al.},
  note = {\url{https://e.huawei.com/eu/material/networking/security
    /0c561b8fd2d342999cd402bcecf6d452}},
  title = {Global DDoS Attack Status 
and Trend Analysis in 2023},
  year = {2024}
}

@inproceedings{hsieh2024netvigil,
  title={$\{$NetVigil$\}$: Robust and $\{$Low-Cost$\}$ Anomaly Detection for $\{$East-West$\}$ Data Center Security},
  author={Hsieh, Kevin and Wong, Mike and Segarra, Santiago and Mani, Sathiya Kumaran and Eberl, Trevor and Panasyuk, Anatoliy and Netravali, Ravi and Chandra, Ranveer and Kandula, Srikanth},
  booktitle={21st USENIX Symposium on Networked Systems Design and Implementation (NSDI 24)},
  pages={1771--1789},
  year={2024}
}

@inproceedings{xu2022xatu,
  title={Xatu: Boosting existing DDoS detection systems using auxiliary signals},
  author={Xu, Zhiying and Ramanathan, Sivaramakrishnan and Rush, Alexander and Mirkovic, Jelena and Yu, Minlan},
  booktitle={Proceedings of the 18th International Conference on emerging Networking EXperiments and Technologies},
  pages={1--17},
  year={2022}
}

@article{cic18-sharafaldin2018toward,
  title={Toward generating a new intrusion detection dataset and intrusion traffic characterization.},
  author={Sharafaldin, Iman and Lashkari, Arash Habibi and Ghorbani, Ali A and others},
  journal={ICISSp},
  volume={1},
  number={2018},
  pages={108--116},
  year={2018}
}

@misc{q1_cloudflare,
  author = {cloudflare},
  note = {\url{https://radar.cloudflare.com/reports/ddos-2025-q1}},
  title = {DDoS threat report for 2025 Q1},
  year = {2025}
}

@inproceedings{alddos_black2022overview,
  title={An overview on detection and prevention of application layer DDoS attacks},
  author={Black, Samuel and Kim, Yoohwan},
  booktitle={2022 IEEE 12th Annual Computing and Communication Workshop and Conference (CCWC)},
  pages={0791--0800},
  year={2022},
  organization={IEEE}
}

@misc{alibaba,
  author = {Alibaba Cloud},
  note = {\url{https://www.alibabacloud.com/help/en/anti-ddos/anti-ddos-pro-and-premium/user-guide/configure-accurate-access-control-rules}},
  title = {Configure the HTTP flood mitigation feature},
  year = {2025}
}

@article{1D-CNN_ige2024state,
  title={State-of-the-art in 1d convolutional neural networks: A survey},
  author={Ige, Ayokunle Olalekan and Sibiya, Malusi},
  journal={IEEE Access},
  year={2024},
  publisher={IEEE}
}

@inproceedings{mawi_kenjiro2000traffic,
  title={Traffic data repository at the WIDE project},
  author={Kenjiro, CHO},
  booktitle={Proc. USENIX 2000 Annual Technical Conference: FREENIX Track, San Diego, CA},
  year={2000}
}

@article{neto2023ciciot2023,
  title={CICIoT2023: A real-time dataset and benchmark for large-scale attacks in IoT environment},
  author={Neto, Euclides Carlos Pinto and Dadkhah, Sajjad and Ferreira, Raphael and Zohourian, Alireza and Lu, Rongxing and Ghorbani, Ali A},
  journal={Sensors},
  volume={23},
  number={13},
  pages={5941},
  year={2023},
  publisher={MDPI}
}

@article{ALDDoS_server2_praseed2018ddos,
  title={DDoS attacks at the application layer: Challenges and research perspectives for safeguarding web applications},
  author={Praseed, Amit and Thilagam, P Santhi},
  journal={IEEE Communications Surveys \& Tutorials},
  volume={21},
  number={1},
  pages={661--685},
  year={2018},
  publisher={IEEE}
}

@article{IPFIX_hofstede2014flow,
  title={Flow monitoring explained: From packet capture to data analysis with netflow and ipfix},
  author={Hofstede, Rick and {\v{C}}eleda, Pavel and Trammell, Brian and Drago, Idilio and Sadre, Ramin and Sperotto, Anna and Pras, Aiko},
  journal={IEEE Communications Surveys \& Tutorials},
  volume={16},
  number={4},
  pages={2037--2064},
  year={2014},
  publisher={IEEE}
}

@article{survey_shen2022machine,
  title={Machine learning-powered encrypted network traffic analysis: A comprehensive survey},
  author={Shen, Meng and Ye, Ke and Liu, Xingtong and Zhu, Liehuang and Kang, Jiawen and Yu, Shui and Li, Qi and Xu, Ke},
  journal={IEEE Communications Surveys \& Tutorials},
  volume={25},
  number={1},
  pages={791--824},
  year={2022},
  publisher={IEEE}
}

@article{DNS_flood_alonso2016mining,
  title={Mining IP to domain name interactions to detect DNS flood attacks on recursive DNS servers},
  author={Alonso, Roberto and Monroy, Ra{\'u}l and Trejo, Luis A},
  journal={Sensors},
  volume={16},
  number={8},
  pages={1311},
  year={2016},
  publisher={MDPI}
}

@article{lititanlog,
  title={TitanLog: Hierarchical and Elastic Logging for High-speed Network Data Stream},
  author={Li, Yuanpeng and Niu, Xian and Zhao, Yikai and Yang, Tong and Hu, Yannan and Zhang, Yuchao and Deng, Xiangwei and Yin, Qiuheng and Zhang, Ziyun and Zhang, Ruwen and others}
}

@inproceedings{li2016flowradar,
  title={$\{$FlowRadar$\}$: A better $\{$NetFlow$\}$ for data centers},
  author={Li, Yuliang and Miao, Rui and Kim, Changhoon and Yu, Minlan},
  booktitle={13th USENIX symposium on networked systems design and implementation (NSDI 16)},
  pages={311--324},
  year={2016}
}

@inproceedings{Securing_Public_Cloud_mani2025securing,
  title={Securing Public Cloud Networks with Efficient Role-based $\{$Micro-Segmentation$\}$},
  author={Mani, Sathiya Kumaran and Hsieh, Kevin and Segarra, Santiago and Chandra, Ranveer and Zhou, Yajie and Kandula, Srikanth},
  booktitle={22nd USENIX Symposium on Networked Systems Design and Implementation (NSDI 25)},
  pages={1033--1048},
  year={2025}
}

@inproceedings{turboflow_sonchack2018turboflow,
  title={Turboflow: Information rich flow record generation on commodity switches},
  author={Sonchack, John and Aviv, Adam J and Keller, Eric and Smith, Jonathan M},
  booktitle={Proceedings of the Thirteenth EuroSys Conference},
  pages={1--16},
  year={2018}
}

@inproceedings{CADE_yang2021cade,
  title={$\{$CADE$\}$: Detecting and explaining concept drift samples for security applications},
  author={Yang, Limin and Guo, Wenbo and Hao, Qingying and Ciptadi, Arridhana and Ahmadzadeh, Ali and Xing, Xinyu and Wang, Gang},
  booktitle={30th USENIX Security Symposium (USENIX Security 21)},
  pages={2327--2344},
  year={2021}
}

@misc{huawei_12000,
  author = {Huawei},
  note = {\url{https://e.huawei.com/au/products/security/12000?from=search}},
  title = {Huawei HiSecEngine AntiDDoS12000 Series DDoS Defense System},
  year = {2026}
}

@article{cv_attribute_xu2020attribute,
  title={Attribute prototype network for zero-shot learning},
  author={Xu, Wenjia and Xian, Yongqin and Wang, Jiuniu and Schiele, Bernt and Akata, Zeynep},
  journal={Advances in Neural Information Processing Systems},
  volume={33},
  pages={21969--21980},
  year={2020}
}

@inproceedings{cv_attribute_al2015transfer,
  title={How to transfer? zero-shot object recognition via hierarchical transfer of semantic attributes},
  author={Al-Halah, Ziad and Stiefelhagen, Rainer},
  booktitle={2015 IEEE winter conference on applications of computer vision},
  pages={837--843},
  year={2015},
  organization={IEEE}
}

@misc{JS-challenge,
  author = {radware},
  note = {\url{https://www.radware.com/cyberpedia/application-security/javascript-challenge/}},
  title = {JavaScript Challenge},
  year = {2026}
}

@article{brute-force_hofstede2017flow,
  title={Flow-based web application brute-force attack and compromise detection},
  author={Hofstede, Rick and Jonker, Mattijs and Sperotto, Anna and Pras, Aiko},
  journal={Journal of network and systems management},
  volume={25},
  number={4},
  pages={735--758},
  year={2017},
  publisher={Springer}
}

@article{rousseeuw1987silhouettes,
  title={Silhouettes: a graphical aid to the interpretation and validation of cluster analysis},
  author={Rousseeuw, Peter J},
  journal={Journal of computational and applied mathematics},
  volume={20},
  pages={53--65},
  year={1987},
  publisher={Elsevier}
}

@misc{netflow_export,
  author = {Netscout},
  note = {\url{https://www.netscout.com/resources/fact-sheets/understanding-netflow-latency-and-effectiveness-in-detecting-and-mitigating-ddos-attacks}},
  title = {Understanding NetFlow Latency and Effectiveness in Detecting and Mitigating DDoS Attacks},
  year = {2026}
}

@inproceedings{patel2025agua,
  title={Agua: A concept-based explainer for learning-enabled systems},
  author={Patel, Sagar and Han, Dongsu and Narodytska, Nina and Jyothi, Sangeetha Abdu},
  booktitle={Proceedings of the ACM SIGCOMM 2025 Conference},
  pages={329--346},
  year={2025}
}

\appendix

\section{Ethical Considerations}

This work does not raise any ethical concerns.

\textbf{User Privacy}. The data used in this paper do not involve user privacy concerns. First, the three datasets used in this study were all legally obtained from their official websites. Second, all real-world traffic traces used in our study were collected with authorization, and sensitive information such as IP addresses was anonymized. Moreover, our detection relies solely on flow data and does not require decrypting user payloads, thereby avoiding the exposure of private content.

\textbf{Attack Isolation.} Although this work studies network traffic detection, it does not cause any harm to real-world networks. All attack generation and traffic replay experiments in this paper were conducted within isolated testbeds.

\section{Robustness Under Flow-Size Drift}

\begin{figure}[h]
    \centering
    \includegraphics[width=0.8\linewidth]{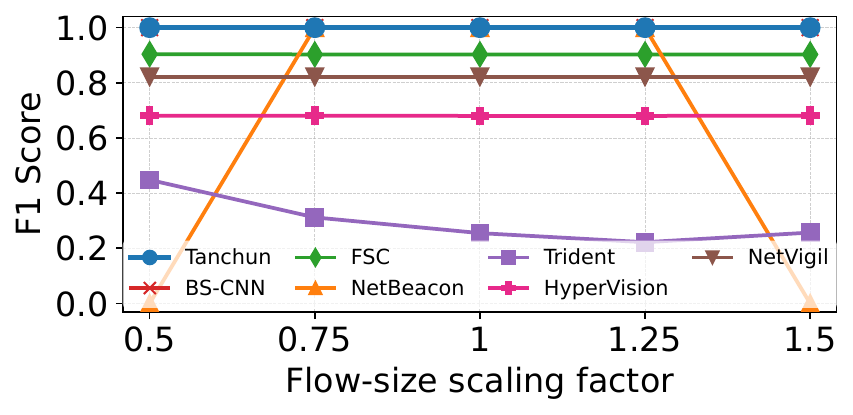}
    \caption{F1-score with varying attack flow size.}
    \label{fig:statistics_drift}
\end{figure}

Attackers can request different legitimate resources to launch HTTP(S) DDoS, causing their flow size to change and challenging the generalization capability of detection models. To simulate this scenario, we use the LOIC trace and proportionally scale the total packet and byte counts of each attack flow to simulate variations in requested resources. Specifically, we train models on the original scale (1x) and evaluate them on the modified datasets with different scaling factors. The results are shown in Fig.~\ref{fig:statistics_drift}.

Tanchun demonstrates strong robustness against flow size drift, maintaining nearly unchanged F1-scores and achieving the best performance. BS-CNN shows similar robustness, as both methods capture IP-pair-level temporal access patterns that remain consistent despite per-flow statistics drift. In contrast, NetBeacon, which relies on single-flow features, fails to detect attacks under large-scale changes (e.g., 0.5$\times$ and 1.5$\times$), as the drifted features deviate from its learned decision boundaries. The anomaly detection methods remain relatively stable since benign traffic distributions are unchanged, but still achieve lower F1-scores than Tanchun.

% \begin{figure}
%     \centering
%     \includegraphics[width=0.8\linewidth]{chart/security_flow_statistics_drift.pdf}
%     \caption{F1-score with varying attack flow statistics.}
%     \label{fig:statistics_drift}
% \end{figure}

\section{Robustness to Attribute Noise}
\label{app_noise}
\begin{figure} [h]
    \centering
    \includegraphics[width=0.8\linewidth]{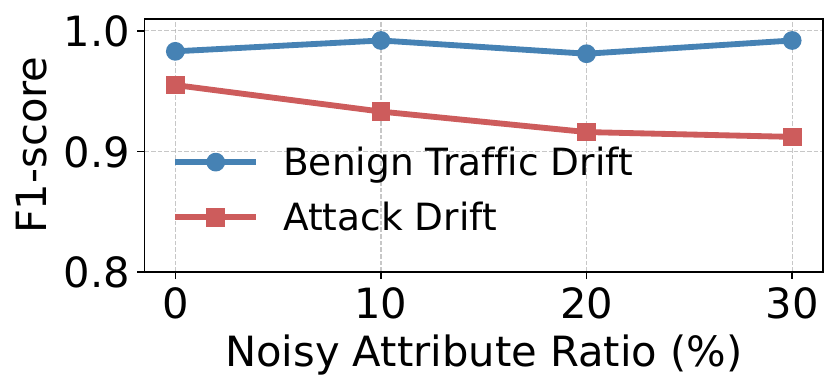}
    \caption{F1-score Variation under Attribute Noise.}
    \label{fig:attr_noisy}
\end{figure}

Since Tanchun relies on behavioral attributes generated from indicators and predefined thresholds, these attributes may contain noise due to imperfect abstraction of real-world behaviors. To evaluate robustness, we randomly perturb the behavioral attributes in the training set by changing the levels of a ratio $R$ of attributes to other levels. We vary $R$ from 0 to 0.3, and the results are shown in Fig.~\ref{fig:attr_noisy}. As the attribute noise increases, Tanchun maintains a nearly unchanged F1-score under benign drift, and the F1-score under attack drift only slightly decreases while remaining above 0.9. These results demonstrate that Tanchun is robust to attribute noise. This is because the original behavior sequences are preserved for detection, and the remaining correct attributes can still provide effective supervision for optimizing the attribute alignment.

\section{Performance Under Cross Attack Events}
\label{whole_zeroshot}

\begin{figure*}
    \centering
    \begin{subfigure}[b]{0.24\linewidth}
        \centering
        \includegraphics[width=0.85\linewidth]{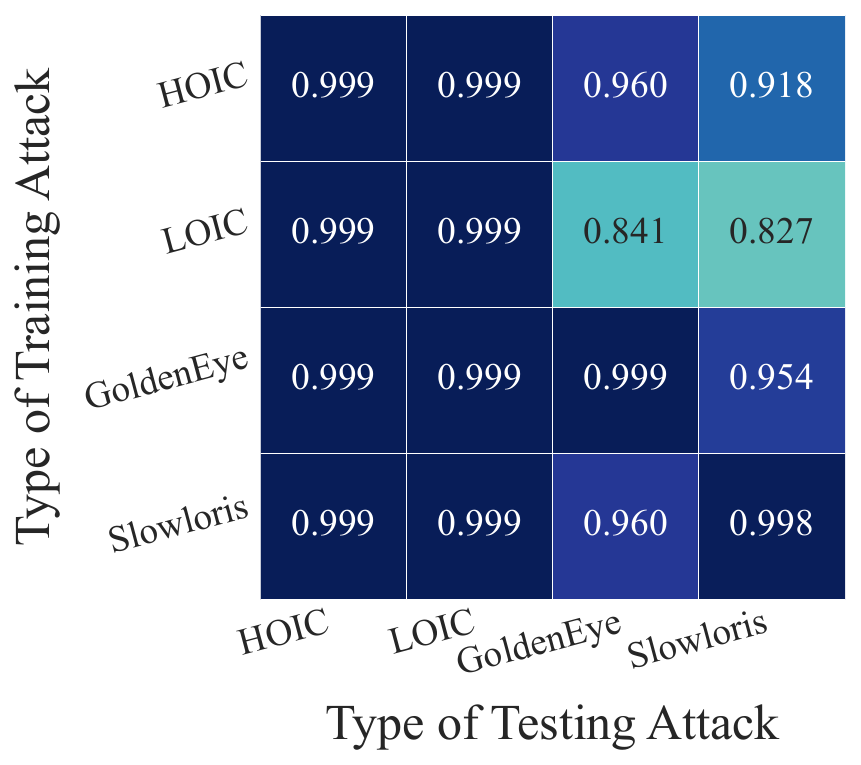}
        \caption{Tanchun}
        \label{fig:tanchun}
    \end{subfigure}
    \hfill
    \begin{subfigure}[b]{0.24\linewidth}
        \centering
        \includegraphics[width=0.85\linewidth]{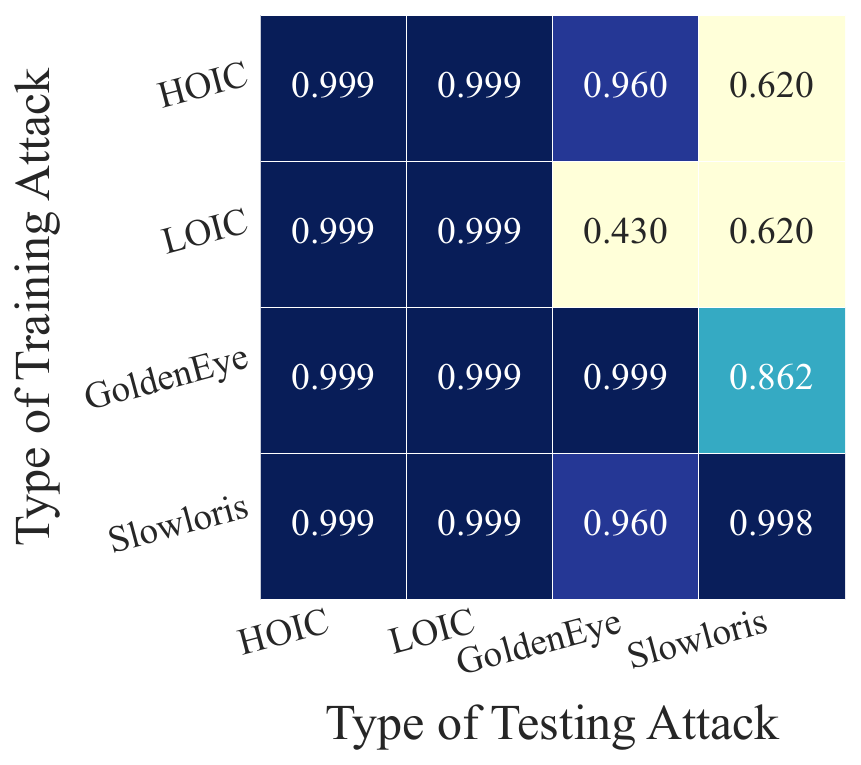}
        \caption{BS-CNN}
        \label{fig:BS-CNN}
    \end{subfigure}
    \hfill
    \begin{subfigure}[b]{0.24\linewidth}
        \centering
        \includegraphics[width=0.85\linewidth]{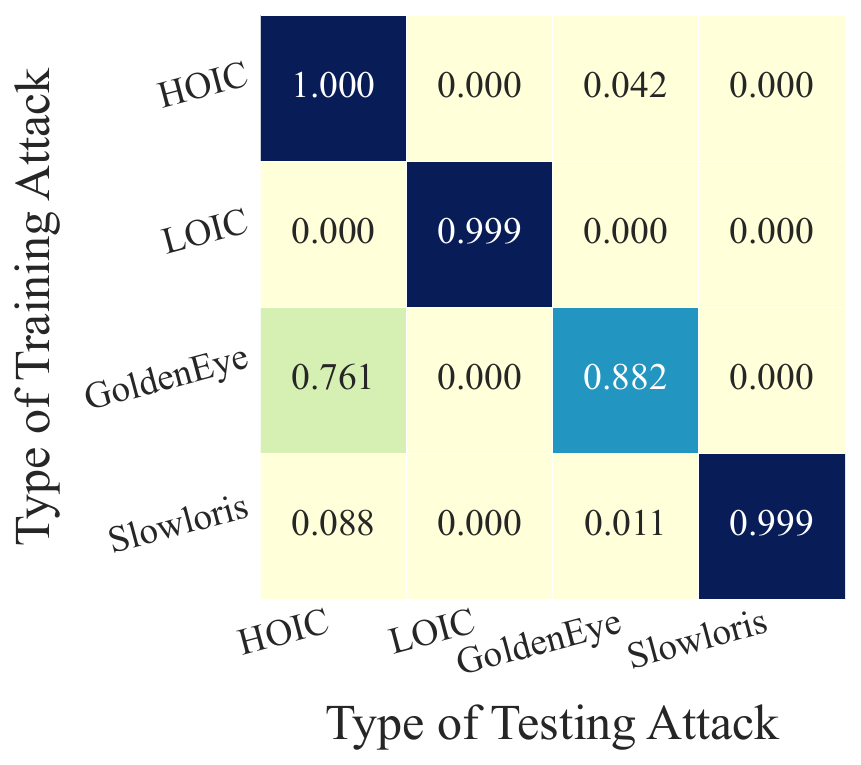}
        \caption{NetBeacon}
        \label{fig:netbeacon}
    \end{subfigure}
    \hfill
    \begin{subfigure}[b]{0.24\linewidth}
        \centering
        \includegraphics[width=0.85\linewidth]{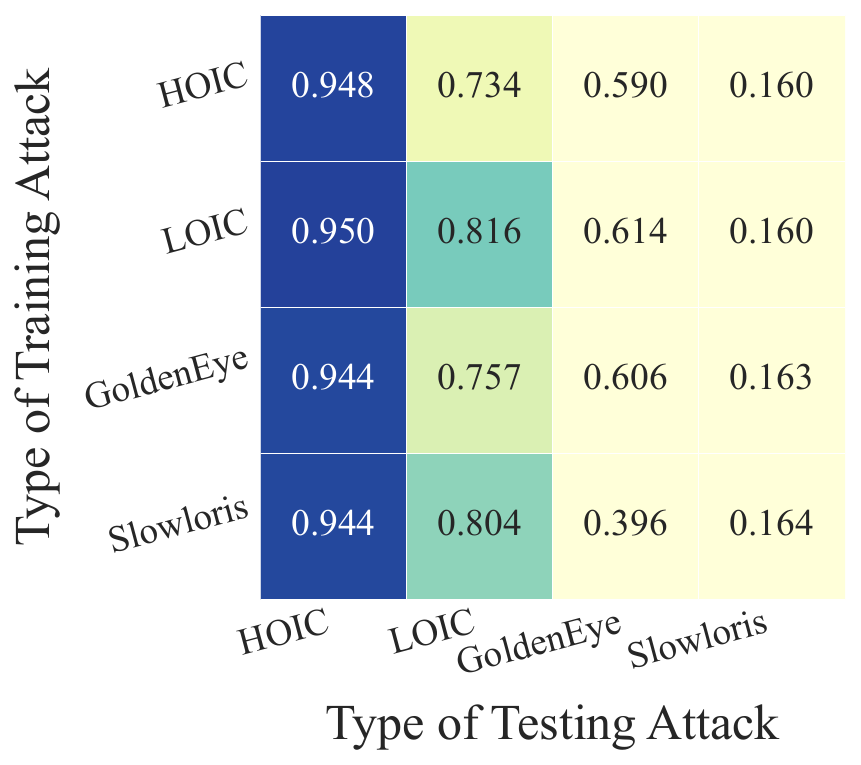}
        \caption{Trident}
        \label{fig:trident}
    \end{subfigure}
    \hfill
    \begin{subfigure}[b]{0.24\linewidth}
        \centering
        \includegraphics[width=0.85\linewidth]{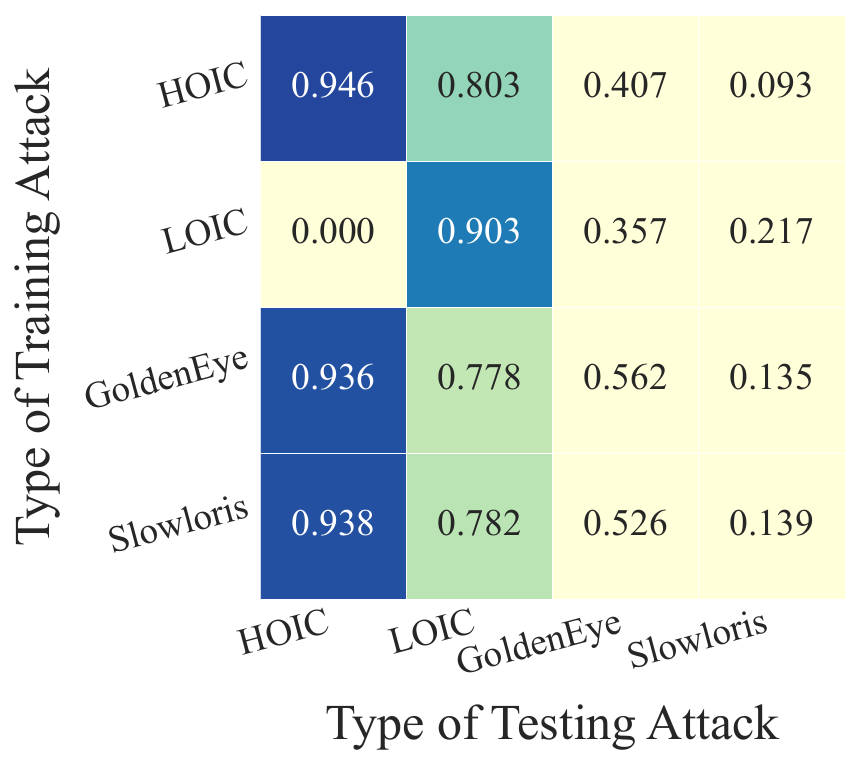}
        \caption{FSC}
        \label{fig:FSC}
    \end{subfigure}
    \hfill
    \begin{subfigure}[b]{0.24\linewidth}
        \centering
        \includegraphics[width=0.85\linewidth]{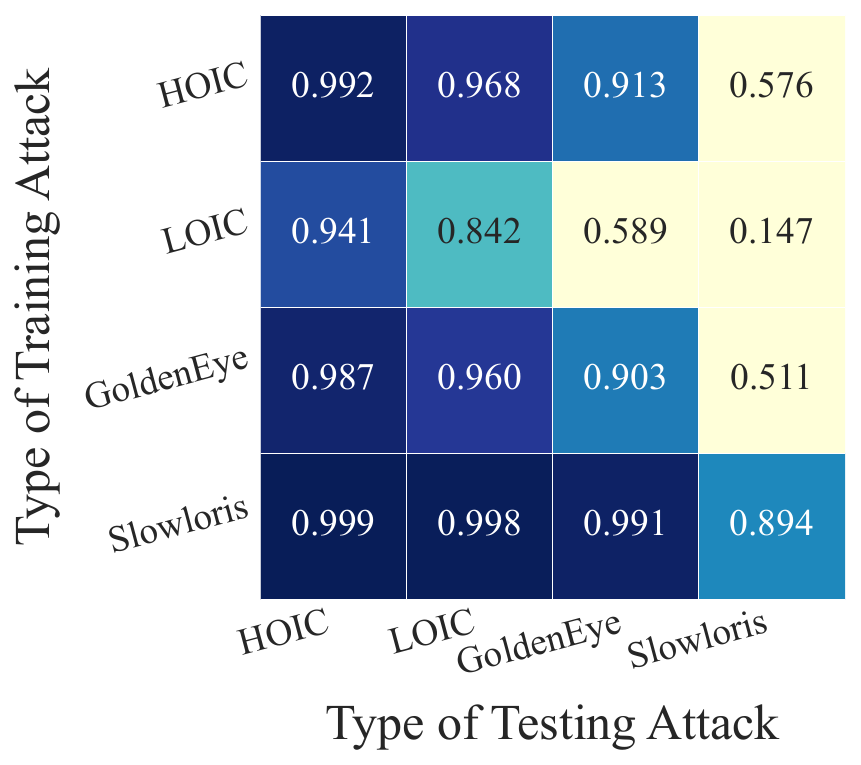}
        \caption{NetVigil}
        \label{fig:netvigil}
    \end{subfigure}
    \hfill
    \begin{subfigure}[b]{0.24\linewidth}
        \centering
        \includegraphics[width=0.85\linewidth]{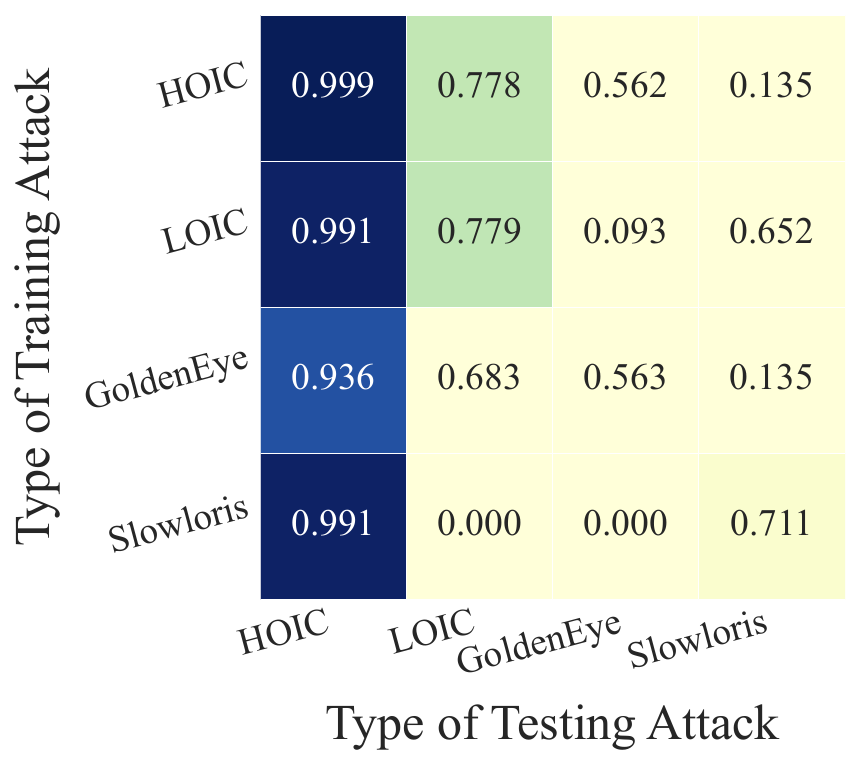}
        \caption{Hypervision}
        \label{fig:netvigil}
    \end{subfigure}

    \caption{F1-score of different methods for unseen-attack detection.}
    \label{fig:whole_zeroshot}
\end{figure*}

We train each model on the training set of each of the four attack traces in CIC-IDS2018 and evaluate it on the test set of all four attacks. The results are shown in Fig.~\ref{fig:whole_zeroshot}. Notably, HyperVision is an outlier-detection method and does not involve explicit training. However, it requires several predefined hyperparameters, which we found to be relatively sensitive. Therefore, for each training set, we search for the best hyperparameter configuration, fix the selected configuration, and use it to evaluate the other attack traces.

%%%%%%%%%%%%%%%%%%%%%%%%%%%%%%%%%%%%%%%%%%%%%%%%%%%%%%%%%%%%%%%%%%%%%%%%%%%%%%%%
\end{document}